\documentclass[journal]{IEEEtran}

\ifCLASSINFOpdf
\else
\fi
\usepackage{nicematrix}
\usepackage{graphicx}
\usepackage{comment}
\usepackage{amsmath}
\usepackage{array}
\usepackage{url}
\usepackage{xcolor}
\usepackage{subcaption}
\usepackage{amssymb}
\usepackage{amsmath}
\usepackage{graphicx}
\usepackage{svg}
\usepackage{array}
\usepackage{booktabs}
\usepackage{slashbox}
\usepackage{hyperref}
\usepackage{orcidlink}
\usepackage{multirow}
\usepackage{multicol}
\usepackage{caption}
\usepackage{boldline}
\usepackage{pifont}
\begin{document}
%
\title{Adaptive Temporal Dynamics for Personalized Emotion Recognition: A Liquid Neural Network Approach}

\author{
Anindya Bhattacharjee\orcidlink{0009-0006-6496-7896}, ~Nittya Ananda Biswas\orcidlink{0009-0009-8608-5714}, ~K. A. Shahriar\orcidlink{0009-0008-2571-7485},
~Adib Rahman\\
Department of Electrical and Electronic Engineering, \\
Bangladesh University of Engineering and Technology
}
\maketitle

\begin{abstract}
Emotion recognition from physiological signals remains challenging due to their non-stationary, noisy, and subject-dependent characteristics. This work presents, to the best of our knowledge, the first comprehensive application of liquid neural networks for EEG-based emotion recognition. The proposed multimodal framework combines convolutional feature extraction, liquid neural networks with learnable time constants, and attention-guided fusion to model temporal EEG dynamics with complementary peripheral physiological and personality features. Dedicated subnetworks are used to process EEG features and auxiliary modalities, and a shared autoencoder-based fusion module is used to learn discriminative latent representations before classification. Subject-dependent experiments conducted on the PhyMER dataset across seven emotional classes achieve an accuracy of 95.45\%, surpassing previously reported results. Furthermore, temporal attention analysis provides interpretable insights into emotion-specific temporal relevance, and t-SNE visualizations demonstrate enhanced class separability, highlighting the effectiveness of the proposed approach. Finally, statistical analysis of temporal dynamics confirms that the network self-organizes into distinct functional groups with specialized fast and slow neurons, proving it independently tunes learnable time constants and memory dominance to effectively capture complex emotion artifacts. 
\end{abstract}


\IEEEpeerreviewmaketitle

\section{Introduction}

The use of physiological signals to recognize someone's emotional state has become very significant in affective computing. Such signals work as physical indicators through the involuntary and continuous responses generated by the autonomic and central nervous systems~\cite{garcia2017emotion}. Unlike facial expressions or vocal cues, which can be deliberately controlled or suppressed, physiological signals such as electroencephalography (EEG), electrocardiography (ECG), electrodermal activity (EDA), and blood volume pulse (BVP) provide objective measures of emotional experiences that are difficult to manipulate. This objectivity makes these physiological modalities valuable for applications requiring authentic emotion assessment, including mental health monitoring, human-computer interaction, driver safety systems, and adaptive learning environments~\cite{picard2000affective}.

Emotions have been modeled both as discrete categories and as continuous dimensions, where complex emotional experiences are described as combinations of more fundamental emotions~\cite{mauss2009measures}. Among dimensional approaches, Russell’s circumplex model is the most widely adopted, representing emotional states along two primary dimensions: valence and arousal~\cite{russell1980circumplex}. This method is the standard in physiological emotion recognition because valence and arousal exhibit distinct patterns in autonomic responses. Arousal strongly correlates with heart rate and skin conductance changes. On the other hand, valence requires more complex integration of central and peripheral signals~\cite{wang2018arousal,alarcao2017emotions,barrett2017theory}. Most benchmark datasets (DEAP~\cite{koelstra2011deap}, DREAMER~\cite{katsigiannis2017dreamer}, AMIGOS~\cite{miranda2018amigos}) adopt this framework to standardize emotional state representation. However, there is no well-defined relationship between these two dimensions. 
Personality traits influence emotional reactivity and physiological regulation, making personality-aware modeling a promising direction for improving subject-independent emotion recognition. 

Current research interests have shifted towards multimodal paradigms that utilize multiple complementary physiological signals for enhanced emotion recognition. Incorporating multiple signals allows proper understanding of the complex nature of emotion experienced by humans. Although EEG signals convey complex information about cortical signal processing, they only address one aspect of emotion. Use cases that incorporate EEG with other peripheral signals address the inherent complementary relationship existing between central and autonomic nervous system responses ~\cite{woo2025deep, lima2024multimodal}. Multimodal signals provide a complementary, diverse set of observations regarding the physiological experience of emotions, with EEG signals resulting from central nervous system processes, while EDA, BVP, and temperature signals result from autonomic activation of specific emotional experiences. This multimodal technique offers two important benefits: each type conveys unique information regarding emotion, with the fact that another type may be impaired due to noise or artifacts. Latest studies show that multimodal models perform emotion classification with higher accuracy and improve robustness regarding sensor failures or signal degradation ~\cite{xu2025emotion, kim2025mifu}.

In this paper, the Physiological Dataset for Multimodal Emotion Recognition (PhyMER)~\cite{pant2023phymer} has been used to facilitate work on personality-aware, subject-dependent emotion recognition. Most benchmarks only assess stimulus–response correlations, with no regard for personality, which is a crucial factor in capturing individual differences in emotional responsiveness. 
PhyMER offers the physiological signals such as EEG, EDA, BVP, and skin temperature. EEG measures the electrical activity produced by synchronized neuronal discharges in the cortical layers, with specific frequency and spatiotemporal profiles clearly linked to particular emotional experiences and cognitive processes~\cite{coan2004frontal}. Such peripheral signals are non-invasive and can be acquired from wearable devices. However, their performance in isolation falls short of that of EEG, and more so in subject-independent scenarios.  


\subsection{Related Works}

Predominantly focused on single-modality approaches, the current trend in emotion recognition is to embrace multimodal solutions. Using multiple sources of information can help to solve the reliability and generalization challenges currently faced in emotion recognition systems. Among different physiological signals, the direct connection with brain activity and good temporal resolution has made EEG the most studied modality.  Bi-hemispheres domain adversarial neural network (BiDANN) by Li et al. achieved 86.15–96.89\% accuracy on ternary classification (positive, negative, neutral) on SEED dataset~\cite{li2018bi}. Cui et al. implemented an end-to-end regional asymmetric CNN network for binary emotion classification (Valence, Arousal) and achieved a 96.65–97.11\% accuracy on the DEAP dataset and a 95.55–97.01\% accuracy on the DREAMER dataset~\cite{cui2020eeg}. Using a convolution and attention-based domain adaptation network, CA2DANet, achieved 95.31\% accuracy on the SEED~\cite{zheng2015investigating} dataset. Their analysis revealed that beta and gamma frequency bands, particularly in frontal and temporal regions, contribute most strongly to emotional discrimination~\cite{cen2025convolution}. Modern approaches to emotion detection from EEG signals employ transformer and attention architectures to capture the temporal and spatial relationships across channels. Utilizing such spectral and temporal attention mechanisms with cross-dataset fine-tuning, Ghous et al. achieved 79- 90\% accuracy across multiple datasets (SEED variants and MPED)~\cite{ghous2025attention}. Attention-based convolutional recurrent neural network was introduced by Tao et al., achieving an accuracy of 93.38–93.72\% on the DEAP dataset and 97.78–98.23\% on the DREAMER dataset~\cite{tao2020eeg}. The network introduced channel-wise attention, CNN for spatial feature extraction, and RNN for temporal information, together to extract discriminative EEG features for different emotions. While these results are impressive, they are obtained under binary or dimensional emotion settings, which significantly simplify the classification problem compared to discrete multi-class emotion recognition. In addition, most of the studies do not model individual emotion states at different timescales. 

Further insights into how the brain regions coordinate during emotional processing can be gained from connectivity analysis. Production of distinctive functional and effective connectivity patterns, particularly in delta and beta bands, by emotional states was shown by Roshanei et al in their EEG-based connectivity patterns during emotional episodes using graph theoretical analysis~\cite{roshanaei2025eeg}.

Among works in peripheral signals, Shu et al. used heart rate features from a smart bracelet to recognize three emotions with 84-96\% accuracy depending on class pairs~\cite{shu2020wearable}, while Khan et al. combined galvanic skin response (GSR) and BVP features to achieve 92\% accuracy using decision trees and k-nearest neighbors~\cite{khan2016recognizing}. However, it should be noted that these examples suffer from limited class distinctions resulting from controlled conditions. 
As such, peripheral signals capture mostly arousal-related autonomic responses while providing limited information about emotion categories, and their utility lies not in replacing EEG but in complementing it.

Effective integration of EEG with peripheral signals requires addressing modality dominance, temporal alignment, and feature-level integration. Woo et al. employed intra- and inter-modality attention mechanisms, along with a proxy-based loss function, in their Modality-Aware Affect Network (MAAN) to address modality imbalance. They obtained 88-92\% accuracy on AMIGOS~\cite{miranda2018amigos} in cross-subject settings by employing contrastive learning to prevent any one modality from dominating the learned representations~\cite{woo2025deep}. Xu et al. presented THHSCA, a hierarchical temporal-spatial fusion framework assessed using the DEAP~\cite{koelstra2011deap}, DREAMER~\cite{katsigiannis2017dreamer}, and PhyMER~\cite{pant2023phymer} benchmark datasets~\cite{xu2025emotion}. Their subject-independent results show the advantages and drawbacks of existing fusion techniques on PhyMER: 80.69\% for valence, 68.90\% for arousal, but only 55.45\% for four-class emotion recognition. The notable drop in multi-class performance shows that despite the model's ability to capture broad dimensional patterns, it is still difficult to differentiate between distinct emotional states across subjects. Using temporal convolutional networks (TCNs) and Modality Quality Indices, Kim et al. developed a quality-aware fusion strategy in the Mifu-ER framework that weights modality contributions according to signal reliability. Performance for seven-class emotion recognition on PhyMER reached 70.24\% with high-quality signals (and 60.55\% when the signal was of inferior quality~\cite{kim2025mifu}. One significant finding from all of these multimodal studies is that the EEG usually shows up as the strongest individual contributor, with peripheral signals offering gradual improvements.

Currently, continuous-time neural models have emerged as a biologically inspired alternative for modeling dynamical systems beyond traditional discrete-time recurrent and attention-based architectures. Neural ordinary differential equation frameworks formalize deep networks as continuous dynamical systems, enabling adaptive temporal evolution \cite{chen2018neural}. Building upon this idea, Liquid Time-Constant (LTC) Networks introduce neurons with learnable time constants, allowing intrinsic adaptation to heterogeneous temporal patterns while maintaining parameter efficiency \cite{hasani2021liquid}. However, their potential for multimodal physiological emotion recognition remains largely unexplored. Despite advances in emotion recognition, existing architectures lack intrinsic mechanisms to adapt to such heterogeneous temporal patterns inherent in physiological signals for different emotions, motivating the exploration of liquid neural networks (LNNs).

\subsection{Research Gap and Our Contribution}
Despite notable advances in physiological emotion recognition, several limitations remain. First, most widely used sequential models, such as LSTMs and GRUs, rely on fixed or implicitly discretized temporal scales, making them ill-suited for multimodal physiological signals that evolve at heterogeneous rates, ranging from rapid EEG fluctuations to slowly varying autonomic responses. Besides, different emotions fluctuate differently in time, and individuals also differ in their emotional variability~\cite{kuppens2013relation}. This mismatch often leads to information loss or inefficient modeling of personalized emotional dynamics. Second, discrete multi-class emotion classification continues to suffer from significant performance degradation, particularly for emotions with similar arousal characteristics. The majority of EEG studies simplify their classification tasks by employing binary (Valence, Arousal) or ternary (Positive, Negative, Neutral) emotional models, and avoids seven-class discrete emotion recognition, leading to inflated accuracy rates~\cite{zhang2024mini} Third, existing multimodal fusion strategies often suffer from modality dominance, where EEG dominates peripheral signals due to inadequate regularization and semantic mismatches between modalities. Finally, many state-of-the-art models lack interpretability and impose high computational overhead, limiting their applicability in real-time and edge-based affective computing systems.

To address these challenges, we propose an attention-augmented LNN framework with learnable time constants for multimodal emotion recognition. By enabling neurons to adapt their temporal dynamics, the model naturally captures both fast and slow emotional processes without explicit supervision. The proposed architecture significantly improves seven-class emotion classification performance on the PhyMER dataset, achieving 95.45\% accuracy with balanced performance across all emotions. We further introduce an autoencoder-based fusion strategy with a regularized bottleneck and annealed reconstruction loss to preserve cross-modal complementarity and prevent modality collapse. Interpretability is enhanced through temporal attention analysis, revealing emotion-specific and physiologically meaningful attention patterns. Finally, the model remains lightweight and efficient, requiring fewer parameters than conventional recurrent architectures while supporting low-latency inference suitable for real-time applications. Together, these contributions establish a principled, interpretable, and deployable solution for multimodal physiological emotion recognition.

\section{Methodology}
\subsection{About the Dataset}

In this work, we specifically focus on the PhyMER dataset~\cite{pant2023phymer}, which consists of EEG, EDA, BVP, and skin temperature data for video-based stimuli, along with the personality traits of 30 participants. Since emotion recognition is highly influenced by individual differences~\cite{zhao2017emotion}, the PhyMER dataset provides a solution to account for these differences by including personality traits. The emotional stimuli in PhyMER were derived from the Korean Video Dataset for Emotion Recognition in the Wild (KVDERW)~\cite{Khanh_Kim_Lee_Yang_Baek_2020}, consisting of video clips designed to elicit seven basic emotions. By bridging personality traits to patterns of physiological responses, PhyMER enables models to capture meaningful individual differences. Data collection involved the use of two distinct wearable devices in which participants viewed 23 stimulus videos, each lasting between 1 and 3 minutes. The sampling rate of EEG, EDA, BVP, and temperature is 256 Hz, 4 Hz, 64 Hz, and 4 Hz, respectively. 

The EEG data has 14 channels, which are AF3, F7, F3, FC5, T7, P7, O1, O2, P8, T8, FC6, F4, F8, AF4. Though the emotional stimuli of each video were labeled by a set of evaluators, some of the videos had poor agreement with the labels annotated by the participants of the experiment. We refer the reader to Table 2 of~\cite{pant2023phymer} for the agreeable percentage of each video. This shows the challenges of a generalized ground truth labeling for emotions, even for humans, further emphasizing the difficulty of emotion detection, which is a highly subject-dependent classification. 

\subsection{Proposed Preprocessing Pipeline}
The preprocessing pipeline of our multimodal model works in parallel on the EEG and other physiological data. The EEG pipeline consists of filtering, independent component analysis (ICA), power spectral density (PSD), differential entropy (DE), and frontal alpha asymmetry (FAA) calculations. For physiological feature extraction, the pipeline also computes HRV metrics from BVP, decomposes EDA into phasic/tonic components, and extracts statistical features from HR and temperature signals.\\

\subsubsection{EEG Signal Conditioning and Artifact Suppression}
The raw EEG signals undergo spectral filtering, blind source separation, power spectral density, and differential entropy calculation, as well as statistical and asymmetrical feature extractions.

Spectral Filtering: EEG signal can contain very low frequency artifacts corresponding to slow drifts such as eye movements and high frequency artifacts corresponding to blinking or muscle activity or even line interference~\cite{uriguen2015eeg}. A zero-phase FIR bandpass filter (1–45 Hz) was applied to remove DC offsets and high-frequency muscle artifacts, followed by notch filters at 50 Hz to eliminate power-line interference.

Blind Source Separation (BSS): Independent component analysis via the FastICA algorithm by MNE was performed to isolate non-cerebral artifacts~\cite{GramfortEtAl2013a}. The core mathematical assumption is that an observed signal, $X$, is a linear mixture of statistically independent source signals, $S$. By finding the demixing matrix, $W$, that can unmix the observed signal to recover the original sources. The process can be modeled by the equation, $X=AS$, where $X$ is the matrix of observed signals and $A$ is the unknown mixing matrix. $S$ is the matrix of independent source signals. 

The key principle is that while the observed signals ($X$) may be Gaussian due to the Central Limit Theorem, the independent sources themselves ($S$) are maximally non-Gaussian. The fastica algorithm iteratively searches for a $W$ matrix that maximizes the non-Gaussianity of the resulting source components. This allows ICA to effectively separate neural signals from high-amplitude, non-Gaussian artifacts like eye blinks and heartbeats~\cite{albera2012ica}.

The ICA first correlates the EEG signal with known EOG artifacts. If it fails, then a heuristic approach is followed to remove noisy artifacts. At first, the variance of each of the independent components is calculated.
\begin{equation}
    V_i = \frac{1}{T-1}\sum_{t=1}^T\left(S_{i,t} - \bar{S}_i\right)^2
\end{equation}
where $S_{i,t}$ is the signal of the $i^{th}$ independent component at time $t$, and $\bar{S}_i$ is the mean of that component's signal. Then the top two components with the highest variance are removed from the EEG data. The heuristic is based on the assumption that strong, dominant artifacts like eye blinks or muscle twitches typically have a much larger signal amplitude than genuine brain activity, leading to a higher variance~\cite{delorme2007enhanced}. The data were re-referenced to the common average, resampled at 128 Hz frequency, and Z-score normalized to mitigate inter-subject impedance variations.
\begin{figure*}
    \centering
    \includegraphics[width=0.9\linewidth]{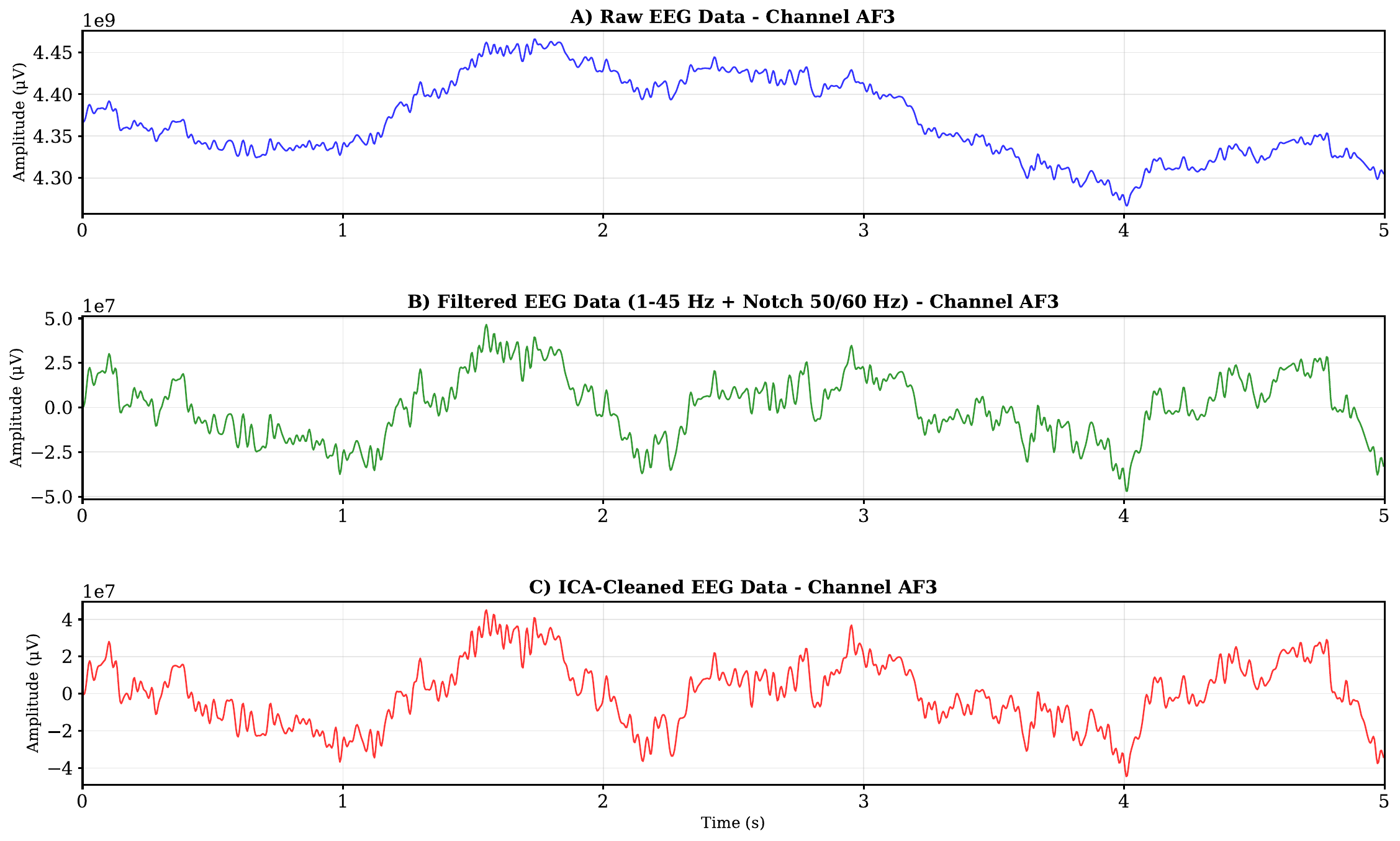}
    \caption{AF3 channel of EEG data for subject 7, video 24 in each step of preprocessing for the first 5 seconds}
    \label{stepwise EEG}
\end{figure*}

Power Spectral Density (PSD): Using Welch’s method~\cite{welch2003use}, we calculated the mean power across five canonical bands: $\delta$ (1-4 Hz), $\theta$ (4-8 Hz), $\alpha$ (8-13 Hz), $\beta$ (13-30 Hz), and $\gamma$ (30-45 Hz). In this process, the input signal, $x[n]$, is divided into $K$ overlapping segments of length $ N$ samples. To prevent spectral leakage caused by abrupt start/end points of the signal, each segment is multiplied by a window function, $w[n]$.

Next, a periodogram is calculated for each windowed segment, which is defined as the squared magnitude of the discrete Fourier transform. 
\begin{equation}
    \hat{P}_m[k] = \frac{1}{LU}\left\lvert\sum_{n=0}^{L-1}x_m[n]w[n]e^{-j2\pi kn/N}\right\rvert^2
\end{equation}

where $U = \frac{1}{L}{\sum_{0}^{L-1}w[n] ^2}$
The final PSD estimate is the average of all the periodograms of the individual segments, $\hat{P}_m[k]$.


Note that a single periodogram is a noisy and inconsistent estimator of the true PSD. By averaging the periodograms, Welch's method significantly reduces the variance of the estimate, producing a much smoother and more reliable PSD plot. In our study, the PSD is calculated for each channel and each 2-second epoch, and then the average power within specific frequency bands is calculated to produce the final features.

Differential Entropy (DE): DE is more temporally stable than raw PSD for affective state mapping~\cite{6695876}. For a continuous random variable $X$, with a probability distribution of $f(X)$, the differential entropy is defined as:
\begin{equation}
    h(x) = -\int f(x)~log(f(x))~dx
\end{equation}

Given that EEG signals often follow a Gaussian distribution, $f(x) = \frac{1}{\sqrt{2\pi \sigma^2}}e^{-\frac{(x - \mu)^2}{2\sigma^2}}$, in specific sub-bands, DE was calculated as $h(x) = \frac{1}{2} \ln(2\pi e \sigma^2)$. 

Statistical Features: Using 4th-order Butterworth bandpass filters constructed in second-order sections, the five canonical frequency bands $\delta$ (1-4 Hz), $\theta$ (4-8 Hz), $\alpha$ (8-13 Hz), $\beta$ (13-30 Hz), and $\gamma$ (30-45 Hz) were initially isolated to extract band-limited statistical information. We have used zero-phase forward-backward filtering \textit{(sosfiltfilt)} to remove phase distortion while preserving temporal alignment of signals. 
If the filter's transfer function is $H(z)$, the effective transfer function $H_{total}(z)$ becomes,
\begin{equation}
  H_{total}(z) = H(z) \cdot H(z^{-1}) = |H(z)|^2
\end{equation}
The phase response of $H(z)$ and $H(z^{-1})$ cancel out exactly, resulting in zero phase distortion. The magnitude response is squared, which doubles the stopband attenuation. Four statistical moments are calculated for each filtered band: mean, standard deviation, skewness, and kurtosis. This produces 280 time-domain features per epoch (5 bands × 4 statistics × 14 channels), which complement frequency-domain measurements by capturing non-linear aspects of emotional brain responses.
\begin{figure*}
    \centering
    \includegraphics[width=0.95\linewidth]{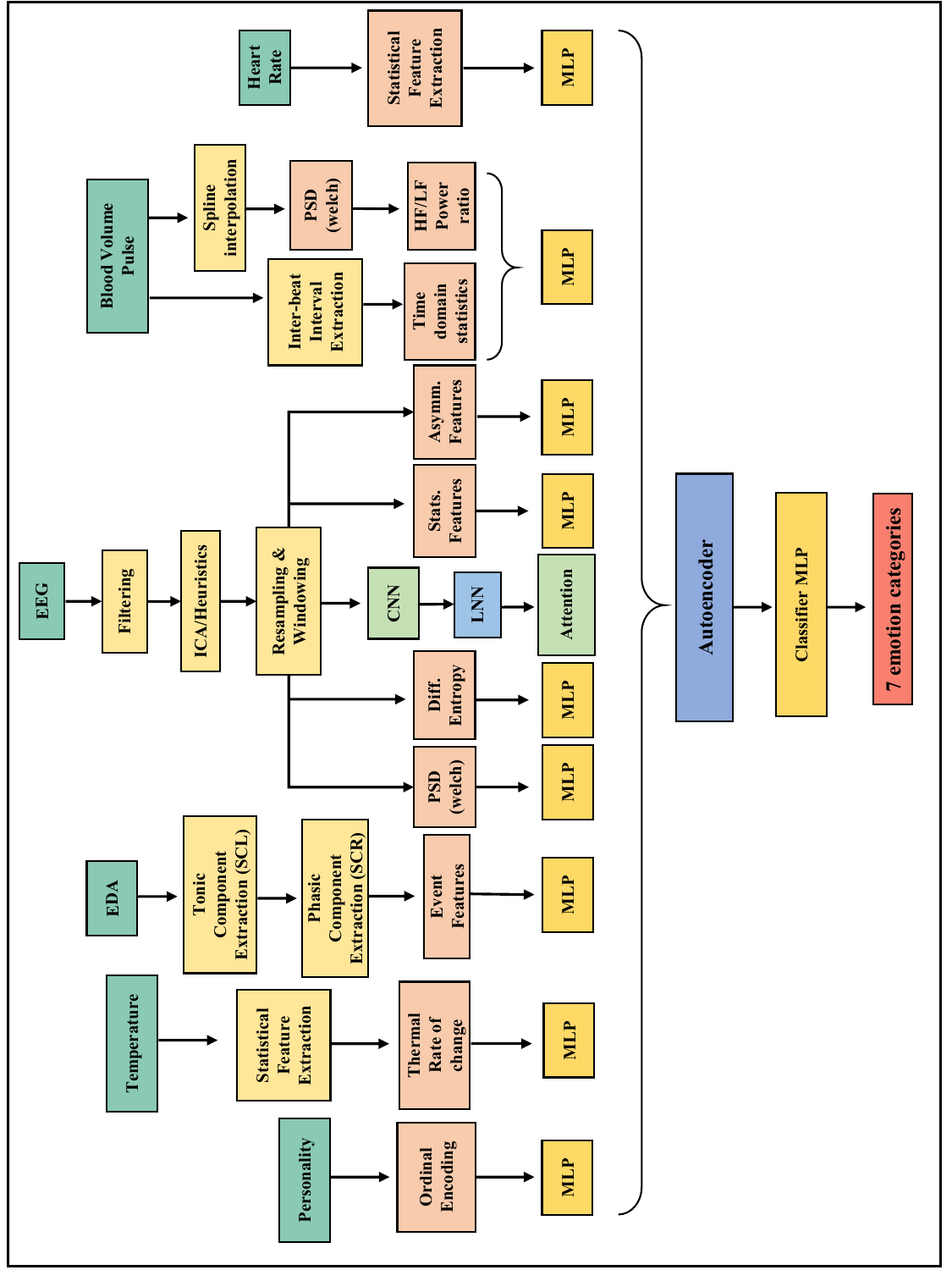}
    \caption{Detailed Preprocessing Pipeline with Architecture}
    \label{fig:placeholder}
\end{figure*}

Frontal Alpha Asymmetry (FAA): We calculated the power differential between right/left frontal pairs (AF4-AF3, F8-F7, F4-F3) in the $\alpha$ band (8-13 Hz) to quantify valence-related approach/withdrawal motivations. 
Under the Approach-Withdrawal model, higher relative left-frontal activity (higher FAA score) is associated with approach motivations and positive emotions like joy or interest. Higher relative right-frontal activity (lower FAA score) is linked to withdrawal motivations and negative emotions like fear, disgust, or sadness~\cite{COAN20047}.\\

\subsubsection{Autonomic Nervous System (ANS) Decomposition}: 
Physiological signals from the Empatica E4 were processed at their native sampling rates to preserve transient features:

Cardiovascular Dynamics (BVP/HR): Inter-beat intervals (IBI) were derived from BVP via peak detection. 
To ensure temporal validity, a minimum inter-peak distance of 500ms or a heart rate of 120 BPM was enforced, combined with a dynamic prominence threshold defined by $0.1\sigma$ of the epoch's local variance. This dual-constraint approach effectively suppressed dicrotic notches and motion artifacts, ensuring high-fidelity IBI extraction. 
After extracting IBI, we calculate the Root Mean Square of Successive Differences (RMSSD), reflecting parasympathetic activity, as well as the  Standard deviation of IBI, representing total HRV. Since IBIs are irregularly sampled, the signal is resampled to a uniform 4 Hz grid before applying the Welch PSD estimate~\cite{welch2003use}. We also calculated the ratio of the Low Frequency ($0.04$–$0.15$ Hz) to the High Frequency ($0.15$–$0.4$ Hz) power, used as an index of sympathovagal balance.

Electrodermal Activity (EDA): The pipeline implements a bimodal decomposition to separate slow-acting and fast-acting sweat gland responses. The tonic component was extracted via a low-pass moving average filter. It represents the basal skin conductance level (SCL), which shifts slowly over minutes. The phasic component was calculated by subtracting the tonic baseline from the raw signal ($S_{phasic} = S_{raw} - S_{tonic}$). This isolates Skin Conductance Responses (SCRs), rapid spikes caused by specific emotional stimuli. The pipeline calculates the SCR Count and the Mean Amplitude within each epoch, which allows for the separation of the slow-varying basal skin conductance from rapid stimulus-evoked responses. 

Heart Rate \& Thermodynamic statistics: Statistical moments like mean, standard deviation, maximum, and minimum value, etc., were extracted from skin temperature and heart rate data to account for peripheral vasoconstriction associated with high-arousal states. Since skin temperature varies slowly, a linear regression is fitted to each epoch: $T(t) = \beta_0 + \beta_1 t$. The slope $\beta_1$ indicates vasoconstriction or vasodilation, which are physiological markers of stress or relaxation. The statistical features of the heart rate provide the model with a coarse-grained view of cardiac arousal compared to the fine-grained HRV data.

\subsubsection{Psychometric Contextualization and Data Structuring}

The pipeline integrates a top-down personality bias into the feature manifold. Each epoch is appended with a vector of the subject’s "Big Five" personality traits (Extraversion, Neuroticism, Agreeableness, Conscientiousness, Openness to Experience), mapped to a 4-point ordinal scale.
Multimodal synchronization was maintained through a rigorous minimum valid epoch check, ensuring that on-time windows with concurrent EEG, BVP, EDA, and TEMP data were included. The final dataset was stratified by emotion label and split (80/20) to ensure the model generalizes across the seven discrete emotion classes.

\subsection{Proposed Model}

In this section, we present the architecture of our proposed hybrid neural network designed for multimodal emotion recognition using physiological signals from the PhyMER dataset. The model integrates CNNs for raw EEG signal processing, LNNs for temporal dynamics modeling, specialized encoders for extracted features from EEG and peripheral signals, an autoencoder for feature fusion, and fully connected layers for final classification. 

The proposed model processes the inputs through modality-specific pathways before fusing them via an autoencoder and classifying emotions into seven categories. Formally, given a set of input modalities $\mathbf{X} = \{\mathbf{X}_{\text{EEG-raw}}, \mathbf{X}_{\text{EEG-PSD}}, \dots, \mathbf{X}_{\text{personality}}\}$, the model computes the feature representations $\mathbf{F} = \{f_m(\mathbf{X}_m) \mid m \in M\}$, concatenates them into a fused vector $\mathbf{F}_{\text{fused}} = [f_1, f_2,\dots,f_{|M|}] \in \mathbb{R}^{D_{fused}}$, encodes them into a latent space via an autoencoder $\phi:\mathbb{R}^{D_{fused}}\to \mathbb{R}^{D_z}$, and outputs the class logits $\mathbf{y} = g(\mathbf{z})$, where $M$ represents the modalities, $\mathbf{z} = \phi(\mathbf{F}_{fused})$ is the latent representation, $f_m$ are modality encoders, and $g$ is the classification head.

The architecture is implemented in PyTorch, with a total parameter count varying by configuration. Dropout regularization~\cite{srivastava2014dropout} is applied uniformly at 0.3 to mitigate overfitting, and Xavier uniform initialization~\cite{glorot2010understanding} is used for weights, with a gain of 0.1 in the final layer to stabilize training. The details of our implemented architecture are presented in Table~\ref{Architecture}.

\begin{table}[htbp]
    \centering
    \caption{Summary of the Model Architecture}
    \label{Architecture}
    \begin{tabular}{l l r}
    \toprule
\textbf{Layer}  & \textbf{Output Shape} & \textbf{\# of params}\\
    \midrule
\textbf{Raw EEG: CNN+LNN+Attention} & & \\
Conv1D + BatchNorm + ReLU & [48, 256] & 4,848 \\
MaxPool + Dropout & [48, 128] & 0 \\
Conv1D + BatchNorm + ReLU & [64, 128] & 21,696 \\
MaxPool + Dropout & [64, 64] & 0 \\
Conv1D + BatchNorm + ReLU & [48, 64] & 21,648 \\
MaxPool + Dropout & [48, 32] & 0 \\
Linear$\times 2$ + Tanh & [128] & 22,656 \\
(LTC Cell + Linear$\times 2$ + Tanh)$\times 32$  & [128] & 128 \\
Linear + Tanh + Dropout & [32, 32] & 4,128 \\
Linear & [32, 1] & 33 \\
\midrule
\textbf{EEG PSD: MLP} & & \\
Linear + ReLU + Dropout & [64] & 4,544 \\
Linear + ReLU + Dropout & [32] & 2,080 \\
\midrule
\textbf{EEG DE: MLP} & & \\
Linear + ReLU + Dropout & [64] & 4,544 \\
Linear + ReLU + Dropout & [32] & 2,080 \\
\midrule
\textbf{EEG Stats: MLP} & & \\
Linear + ReLU + Dropout & [128] & 35,968 \\
Linear + ReLU + Dropout & [64] & 8,256 \\
Linear + ReLU + Dropout & [32] & 2,080 \\
\midrule
\textbf{EEG Asymmetry: MLP} & & \\
Linear + ReLU + Dropout & [16] & 64 \\
Linear + ReLU & [8] & 136 \\
\midrule
\textbf{Heart Rate Variability: MLP} & & \\
Linear + ReLU + Dropout & [32] & 256 \\
Linear + ReLU & [16] & 528 \\
\midrule
\textbf{Electrodermal Activity: MLP} & & \\
Linear + ReLU + Dropout & [32] & 288 \\
Linear + ReLU & [16] & 528 \\
\midrule
\textbf{Heart Rate: MLP} & & \\
Linear + ReLU + Dropout & [32] & 256 \\
Linear + ReLU & [16] & 528 \\
\midrule
\textbf{Temperature: MLP} & & \\
Linear + ReLU + Dropout & [32] & 224 \\
Linear + ReLU & [16] & 528 \\
\midrule
\textbf{Personality: MLP} & & \\
Linear + ReLU + Dropout & [32] & 192 \\
Linear + ReLU & [16] & 528 \\
\midrule
\textbf{Fusion Layer: Autoencoder} & & \\
Linear + ReLU + Dropout & [256] & 80,128\\
Linear + ReLU + Dropout & [128] & 32,896\\
Linear + ReLU + Dropout & [256] & 33,024\\
Linear & [312] & 80,184\\
Autoencoder & [128] & 0\\
\midrule
\textbf{Classification: MLP} & & \\
Linear+BatchNorm+ReLU+Dropout & [256] & 33,536\\
Linear+BatchNorm+ReLU+Dropout & [128] & 33,152\\
Linear  & [7] & 903\\
\midrule
\textbf{Total trainable parameters:} & & 432,568\\ 
    \bottomrule
    \end{tabular}
\end{table}

\subsubsection{Processing Raw EEG: CNN-LNN-Attention}
Raw EEG signals $\mathbf{X}_{\text{EEG-raw}} \in \mathbb{R}^{B \times C \times T} $ (where $B$ is batch size, $C=14$ channels, and $T=256$ timesteps) are first processed by a CNN to extract spatial features, followed by an LNN to model temporal dynamics. The CNN consists of three convolutional blocks with filter sizes $[n_1, n_2, n_3]=[48, 64, 48]$. Each block applies:
\begin{equation}
    h^{(l)} = MP(D(\text{ReLU}(BN(W^{l}*h^{(l-1)}+b^l))), p = 0.3)
\end{equation}
Here $*$ denotes 1D convolution with kernel size $k=7$ and padding $p=3$, $MP$ is max pooling, $D$ is drouput and $BN$ is batch normalization, and $h^{(0) }=\mathbf{X}_{\text{EEG-raw}}$. Max-pooling with stride 2 progressively reduces the temporal dimension, yielding $\mathbf{H}_{CNN} \in \mathbb{R}^{B\times n_3\times T'}$ where $T' = [T/2^3] =32$. 

The CNN output is transposed to $\mathbf{H}_{CNN}^T \in \mathbb{R}^{B\times T' \times n_3}$ and fed into a multi-layer LNN. Unlike traditional RNNs or LSTMs that operate on fixed time scales with static gates, the LNN model utilizes an LTC cell~\cite {hasani2021liquid} that adapt their temporal dynamics with learnable time constants. The continuous-time formulation of an LTC neuron is governed by: 
\begin{equation}
    \tau \frac{dh(t)}{dt} = -h(t)+\sigma(W_xx(t)+W_hh(t)+b)
\end{equation}
For computational efficiency, we discretize this using the exponential integrator method with time step $\Delta t$:
\begin{equation}
\mathbf{d} = \exp\left(-\frac{\Delta t}{\boldsymbol{\tau}}\right)
\end{equation}

\begin{equation}
\mathbf{h}_t = \mathbf{d} \odot \mathbf{h}_{t-1} + (1 - \mathbf{d}) \odot \tanh(W_x \mathbf{x}_t + W_h \mathbf{h}_{t-1} + b)
\end{equation}
where $\odot$ denotes element-wise multiplication, $\boldsymbol{\tau} \in \mathbb{R}^{d_h}$ is a vector of learnable time constants (one per hidden unit), and $\mathbf{d} \in \mathbb{R}^{d_h}$ is the decay factor. Note that, $\boldsymbol{\tau}$ is parameterized in log-space as $\boldsymbol{\tau} = \exp(\boldsymbol{\theta}_{\tau})$
to enforce positivity, where $\boldsymbol{\theta}_{\tau}$ are the actual learned parameters initialized uniformly in $[\log(0.1), \log(10)]$, corresponding to time constants in [0.1,10]. This formulation provides several advantages. Firstly, each hidden unit learns its own memory timescale, enabling the network to simultaneously capture both rapid transient responses (small $\tau$) and slow emotional state changes (large $\tau$) in EEG signals. Secondly, the model can theoretically handle irregularly sampled data, allowing continuous time data, though our implementation uses uniform $\Delta t = 1$. Lastly, the exponential decay ensures $\mathbf{h}_t$ remains bounded regardless of sequence length.

For multi-layer LNNs with $L$ layers, the hidden state at layer $l$ and time $t$ is computed as:
\begin{equation}
\begin{split}
    \mathbf{h}_t^{(l)} &= \mathbf{d}^{(l)} \odot \mathbf{h}_{t-1}^{(l)} + (1 -\mathbf{d}^{(l)}) \odot \tanh(W_x^{(l)} \mathbf{h}_t^{(l-1)} \\
    &+ W_h^{(l)} \mathbf{h}_{t-1}^{(l)} + b^{(l)}) 
\end{split}
\end{equation}

where $\mathbf{h}_t^{(0)} = \mathbf{x}_t$ is the input at time $t$. Dropout with rate $p=0.3$ is applied between layers during training. The LNN produces a sequence of hidden states $\mathbf{H}_{\text{LNN}} = [\mathbf{h}_1^{(L)}, \mathbf{h}_2^{(L)}, \ldots, \mathbf{h}_{T'}^{(L)}] \in \mathbb{R}^{B \times T' \times d_h}$ where $d_h = 128$ is the hidden dimension.

Rather than using only the final hidden state $\mathbf{h}_{T'}^{(L)}$, we employ a self-attention mechanism to aggregate the entire sequence. The attention weights are computed as:
\begin{equation}
\mathbf{e}_t = \mathbf{v}^T \tanh(W_a \mathbf{h}_t^{(L)} + b_a)
\end{equation}
\begin{equation}
\alpha_t = \frac{\exp(\mathbf{e}_t)}{\sum_{t'=1}^{T'} \exp(\mathbf{e}_{t'})}
\end{equation}
\begin{equation}
\mathbf{f}_{\text{EEG-raw}} = \sum_{t=1}^{T'} \alpha_t \mathbf{h}_t^{(L)}
\end{equation}

where $W_a \in \mathbb{R}^{d_a \times d_h}$ with $d_a = 32$ is the attention projection matrix, $\mathbf{v} \in \mathbb{R}^{d_a}$ is a learnable context vector, and ${\alpha} = [\alpha_1, \alpha_2, \dots, \alpha_{T'}]$ are the normalized attention scores. This mechanism allows the model to focus on emotionally salient temporal segments within the EEG epoch, yielding $\mathbf{f}_{\text{EEG-raw}} \in \mathbb{R}^{B \times d_h}$.

\subsubsection{Feature Encoders for EEG \& Peripheral Signals}: 
In addition to raw EEG processing, the model incorporates specialized encoders for pre-extracted features from multiple modalities. Each encoder $f_m$ is a fully connected network tailored to the dimensionality and characteristics of its input modality.

PSD features $\mathbf{X}_{\text{PSD}} \in \mathbb{R}^{B \times 14 \times 5}$, and DE features $\mathbf{X}_{\text{DE}} \in \mathbb{R}^{B \times 14 \times 5}$ are flattened and processed through two-layer networks with dimensions $70 \rightarrow 64 \rightarrow 32$, and similarly for DE features.
\begin{equation}
    \mathbf{f}_{\text{PSD}} = \text{ReLU}(W_2 \cdot \text{D}(\text{ReLU}(W_1 \cdot \mathbf{X}_{\text{PSD}} + b_1)) + b_2)
\end{equation}
Here $D$ denotes dropout,t which is implemented with a probability of 0.3.
For the high-dimensional statistical features $\mathbf{X}_{\text{stats}} \in \mathbb{R}^{B \times 14 \times 20}$, a deeper three-layer encoder is used with dimensions $\rightarrow 128 \rightarrow 64 \rightarrow 32$, providing sufficient capacity to capture complex non-linear relationships among statistical measures. Frontal alpha asymmetry features $\mathbf{X}_{\text{asym}} \in \mathbb{R}^{B \times 3}$ are encoded via a compact two-layer network ($\rightarrow 16 \rightarrow 8$), reflecting the lower dimensionality of asymmetry features. HRV (7 features), EDA (8 features), Heart Rate statistics (7 features), and skin Temperature (6 features) are each processed through two-layer encoders with architecture $d_{\text{in}} \rightarrow 32 \rightarrow 16$, where $d_{\text{in}}$ is modality-specific. The Big Five personality scores $\mathbf{X}_{\text{personality}} \in \mathbb{R}^{B \times 5}$
 are also encoded via a two-layer network (5 $\rightarrow 32 \rightarrow 16$), allowing the model to condition emotional responses on stable individual differences.

\subsubsection{Autoencoder-based feature fusion}

The encoded features from all modalities are concatenated to form:
$\mathbf{F}_{\text{fused}} = [\mathbf{f}_{\text{EEG-raw}}; \mathbf{f}_{\text{PSD}}; \mathbf{f}_{\text{DE}}; \mathbf{f}_{\text{stats}}; \mathbf{f}_{\text{asym}}; \mathbf{f}_{\text{HRV}}; \mathbf{f}_{\text{EDA}}; \mathbf{f}_{\text{HR}}; \mathbf{f}_{\text{TEMP}}; \mathbf{f}_{\text{personality}}]$
where $\mathbf{F}_{\text{fused}} \in \mathbb{R}^{B \times D_{\text{fused}}}$
with $D_{\text{fused}} = 128 + 32 + 32 + 32 + 8 + 16 + 16 + 16 + 16 + 16 = 312$ when all modalities are used. To learn a compact and robust latent representation, we employ an autoencoder with encoder $\phi_{\text{en}}: \mathbb{R}^{D_{\text{fused}}} \rightarrow \mathbb{R}^{D_z}$ and decoder $\phi_{\text{de}}: \mathbb{R}^{D_z} \rightarrow \mathbb{R}^{D_{\text{fused}}}$.

\begin{equation}
\begin{split}
\mathbf{z} &= \phi_{\text{en}}(\mathbf{F}_{\text{fused}}) \\
&= \text{ReLU}(W_e^2 \cdot \text{D}(\text{ReLU}(W_e^1 \cdot \mathbf{F}_{\text{fused}} + b_e^2)) + b_e^1)
\end{split}
\end{equation}

\begin{equation}
\begin{split}
\hat{\mathbf{F}}_{\text{fused}} &= \phi_{\text{dec}}(\mathbf{z})\\
&= W_d^2 \cdot \text{D}(\text{ReLU}(W_d^1 \cdot \mathbf{z} + b_d^1)) + b_d^2
\end{split}
\end{equation}

With bottleneck dimension $D_z = 128$, creating an information bottleneck that forces the model to learn salient cross-modal patterns. The autoencoder is trained jointly with the classifier using a reconstruction loss:
\begin{equation}
\mathcal{L}_{\text{recon}} = \|\mathbf{F}_{\text{fused}} - \hat{\mathbf{F}}_{\text{fused}}\|_2^2    
\end{equation}

This regularization encourages the latent space $\mathbf{z}$ to preserve modality complementarity while discarding noise, a critical property for multimodal fusion.

\subsubsection{Classification Head}
The latent representation $\mathbf{z}$ is passed through a three-layer fully connected classification network with dimensions 1$28 \rightarrow 256 \rightarrow 128 \rightarrow 7$, where 7 corresponds to the emotion classes. Batch normalization is applied after the first two layers to stabilize training. The final layer weights $\mathbf{W}_3$ are initialized with Xavier uniform initialization with gain 0.1 to prevent large initial logits that can lead to premature confidence and mode collapse during early training.

The model is trained end-to-end by minimizing a composite loss function:
\begin{equation}
\mathcal{L} = \mathcal{L}_{\text{CE}}(\mathbf{y}, \mathbf{y}_{\text{true}}) + \lambda(e) \cdot \mathcal{L}_{\text{recon}}
\end{equation}
where $\mathcal{L}_{\text{CE}}$ is the cross-entropy loss with label smoothing ($\epsilon = 0.1$) and class weights to handle class imbalance. Note that class weights for handling imbalance are as follows: angry: 1.337, disgust: 1.124, fear: 1.149, happy: 1.115, neutral: 0.809, sad: 0.591, surprise: 1.501. $\mathcal{L}_{\text{recon}}$ is the reconstruction loss defined above, and $\lambda(e) = \lambda_0 (1 - e/E)$ is an epoch-dependent weight that anneals the reconstruction term from $\lambda_0 = 0.001$ to 0 over $E$ epochs. This annealing strategy prioritizes reconstruction early in training to learn robust multimodal features, then gradually shifts focus to classification performance.

\section{Results}

\begin{figure*}[t]
    \centering
    \begin{subfigure}[b]{0.56\textwidth}
        \centering
        \includegraphics[width=\linewidth]{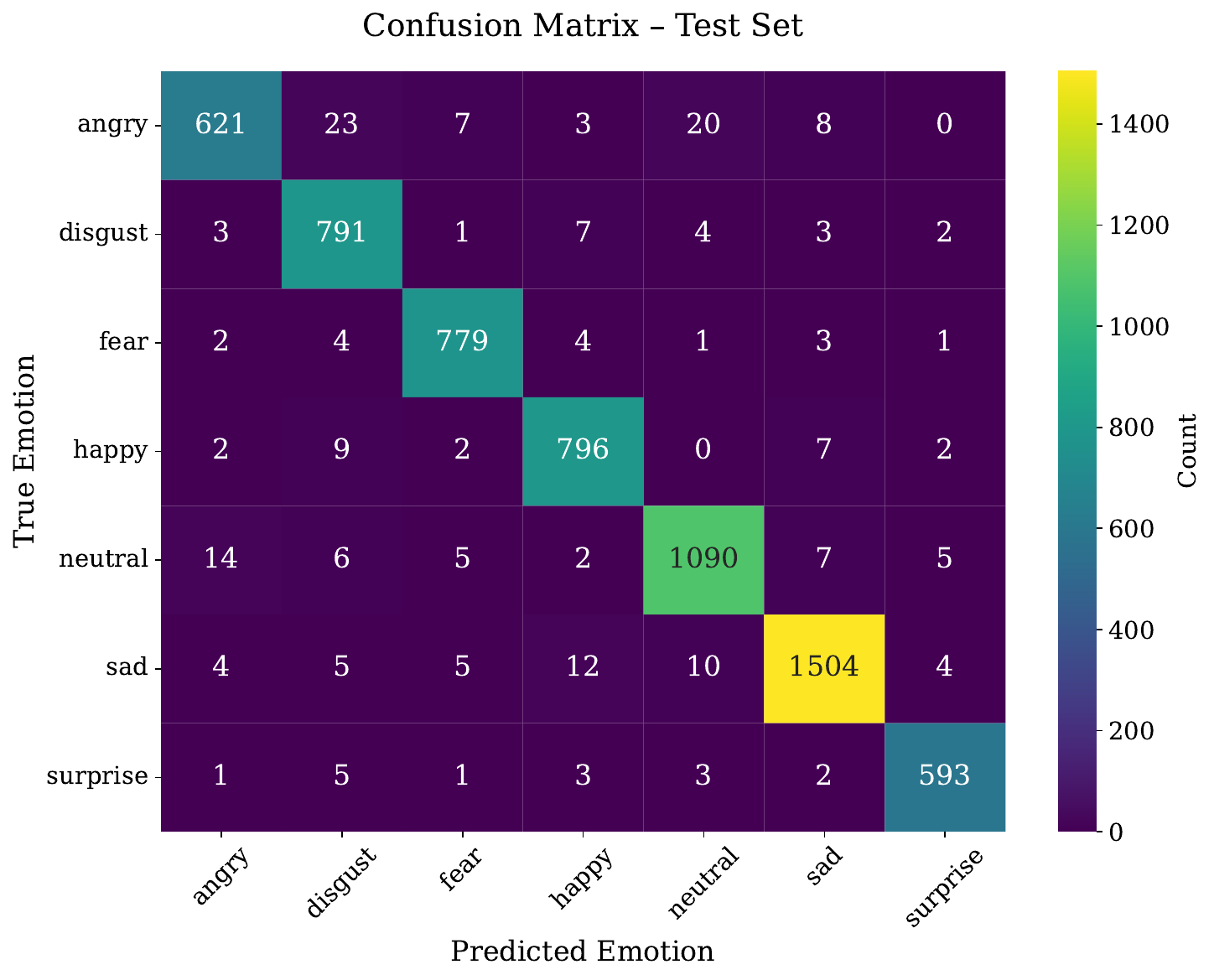}
        \caption{}
        \label{fig:confusion matrix}
    \end{subfigure}
    \hfill
    \begin{subfigure}[b]{0.43\textwidth}
        \centering
        \includegraphics[width=\linewidth]{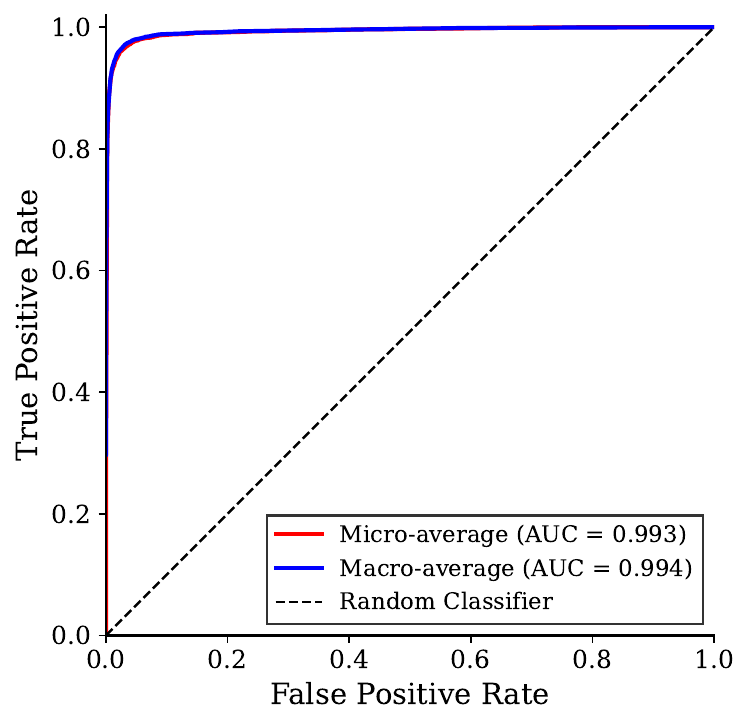}
        \caption{}
        \label{fig:roc_curves}
    \end{subfigure}
    \caption{Classification performance analysis: (a) Confusion matrix for test set, (b) Micro and macro-average ROC curves.}
    \label{fig:classification_analysis}
\end{figure*}

In this section, we present the performance of our model on the PhyMER dataset. Table~\ref{tab:classification_metrics} shows the overall performance of our proposed network for subject-dependent analysis. The mean and the standard deviation of the 7 different metrics are calculated over five different random seeds. Note that the accuracy is recorded at 95.45\% with a balanced accuracy of 94.63\%, which shows excellent classification accuracy. 
Cohen’s $\kappa$ value of 0.9338 shows near-perfect agreement between the actual labels and the predicted labels by the model. Similarly, the Matthews Correlation Coefficient of 0.9437 confirms strong predictive performance~\cite{chicco2023matthews}. The low log loss value of 0.2634 indicates well-calibrated probability estimates.
\begin{table}[htbp]
    \centering
    \caption{Classification metrics with all modalities for subject-dependent analysis}
    \label{tab:classification_metrics}
    \begin{tabular}{l c}
        \toprule
        \textbf{Metric} & \textbf{Value} \\
        \midrule
        Accuracy & 0.9545 $\pm$ 0.014   \\
        Balanced Accuracy & 0.9463 $\pm$ 0.011  \\
        Macro F1-Score & 0.9371 $\pm$ 0.019 \\
        Weighted F1-Score & 0.9489 $\pm$ 0.016 \\
        Cohen's Kappa & 0.9338 $\pm$ 0.023 \\
        Matthews Correlation & 0.9437 $\pm$ 0.015 \\
        Log Loss & 0.2634 $\pm$ 0.013 \\
        \bottomrule
    \end{tabular}
\end{table}

We present a detailed analysis in Table~\ref{tab:per_class_analysis} to investigate the classification performance of individual emotions. Note that the system is capable of reliably classifying between all seven emotion categories despite having a significant class imbalance. The AUC-ROC scores exceed 0.98 for every class, with a macro-average of 0.99, demonstrating near-perfect discriminative ability. This can also be seen from Fig.~\ref{fig:roc_curves}, where both micro-average (AUC = 0.993) and macro-average (AUC = 0.994) approaches yield near-identical results, indicating consistent discrimination across all classes regardless of sample frequency. The curves' proximity to the upper-left corner confirms minimal trade-off between sensitivity and specificity. 

\begin{table}[htbp]
    \centering
    \caption{Classification metrics for individual classes for subject-dependent analysis}
    \label{tab:per_class_analysis}
    \begin{tabular}{l c c r r r r}
        \toprule
        \textbf{Emotion} & \textbf{Support} & \textbf{\%} & \textbf{Precision} & \textbf{Recall} & \textbf{F1} & \textbf{AUC} \\
        \midrule
        Angry & 682 & 10.7 & 0.95 & 0.83 & 0.88 & 0.98 \\
        Disgust & 811 & 12.7 & 0.92 & 0.96 & 0.94 & 0.99 \\
        Fear & 794 & 12.4 & 0.98 & 0.92 & 0.95 & 0.99 \\
        Happy & 818 & 12.8 & 0.96 & 0.97 & 0.96 & 0.99 \\
        Neutral & 1129 & 17.7 & 0.90 & 0.97 & 0.93 & 0.99 \\
        Sad & 1544 & 24.2 & 0.94 & 0.96 & 0.95 & 0.99 \\
        Surprise & 608 & 9.5 & 0.97 & 0.94 & 0.95 & 0.99 \\
        \midrule
        \textbf{Average} & -- & -- & \textbf{0.946} & \textbf{0.936} & \textbf{0.937} & \textbf{0.989} \\
        \bottomrule
    \end{tabular}
\end{table}

We present the top 10 misclassification patterns to better understand the model's bottlenecks in Table~\ref{tab:misclassifications}. The most frequent confusion occurs between 'Angry' and 'Disgust' at 3.37\%, followed by 'Angry' and 'Neutral' at 2.93\%. These patterns can be understood because negative valence emotions like 'Angry' and 'Disgust' have similar autonomic responses ~\cite{schechtman2010negative}. Again,  these misclassifications show that 'Angry' is often misclassified as another class, but not the other way around. This one-sided nature suggests that the physical signs of anger may be more ambiguous compared to other classes, which can also be seen by the low recall of 'Angry' presented in Table~\ref{tab:per_class_analysis}. Confusion between 'Angry' and 'Sad' at 1.17\% also reflects common negative valence emotional traits. A detailed analysis of ambiguity across different emotions can be identified from the confusion matrix presented in Figure~\ref{fig:confusion matrix}. 

\begin{table}[htbp]
    \centering
    \caption{Top 10 misclassification patterns}
    \label{tab:misclassifications}
    \begin{tabular}{l l r r}
        \toprule
        \textbf{True Class} & \textbf{Predicted As} & \textbf{Count} & \textbf{Rate} \\
        \midrule
        Angry & Disgust & 23 & 3.37\% \\
        Angry & Neutral & 20 & 2.93\% \\
        Neutral & Angry & 14 & 1.24\% \\
        Sad & Happy & 12 & 0.07\% \\
        Sad & Neutral & 10 & 0.06\% \\
        Happy & Disgust & 9 & 1.10\% \\
        Angry & Sad & 8 & 1.17\% \\
        Angry & Fear & 7 & 1.02\% \\
        Neutral & Sad & 7 & 0.06\% \\
        Happy & Sad & 7 & 0.08\% \\
        \bottomrule
    \end{tabular}
\end{table}

\subsection{Choice of Hyperparameters}

Due to computational resource constraints, a comprehensive hyperparameter tuning using advanced optimization algorithms was not feasible. Instead, we conducted a discrete evaluation of key hyperparameters to identify optimal settings. We assessed time constant ranges (0.1, 10), (0.1, 15), and (0.1, 20),(each with a fixed learning rate of $5\times10^{-4}$ and batch size of 64); learning rates of $1\times10^{-3}$, $5\times10^{-4}$, and $1\times10^{-4}$ (each with a fixed time constant range (0.1, 10) and batch size of 64) and batch sizes of 16, 32, 64 (each with a fixed learning rate of $5\times10^{-4}$ and time constant range (0.1, 10)). The results are presented in Table~\ref{tab:hyperparameter}. We acknowledge this as a limitation of our work and plan to incorporate a more comprehensive hyperparameter study in the future. The results show that the optimum performance is achieved for a time constant ranging from 0.1 to 10, with a learning rate of $5\times10^{-4}$ and a batch size of 64.

\begin{table}[htbp]
    \centering
    \caption{Accuracy results for different hyperparameter configurations}
    \label{tab:hyperparameter}
    \begin{tabular}{l l}
    \toprule
     Parameter & Accuracy (\%) \\
     \midrule
    Time constant range: & \\
     ~~~~(0.1, 10) & 95.45 $\pm$ 1.40\\
     ~~~~(0.1, 15) & 94.11 $\pm$ 1.13\\
     ~~~~(0.1, 20) & 94.21 $\pm$ 0.92\\
    \hline
     Learning rate: & \\
     ~~~~$1\times10^{-3}$ & 95.41 $\pm$ 1.28\\
     ~~~~$5\times10^{-4}$ & 95.45 $\pm$ 1.40\\
     ~~~~$1\times10^{-4}$ & 95.38 $\pm$ 1.76\\
    \hline
     Batch size: & \\
     ~~~~~~16 & 94.17 $\pm$ 1.25\\
     ~~~~~~32 & 94.91 $\pm$ 1.72 \\
     ~~~~~~64 & 95.45 $\pm$ 1.40\\
     \bottomrule
    \end{tabular}

\end{table}

\begin{table*}[htbp]
    \centering
    \caption{Ablation study on the model architecture with all modalities considered}
    \label{Architecture ablation}
    \begin{NiceTabular}{c c c c c c | c c}
    \toprule
ID & Liquid Hidden & \# of Liquid & CNN Filters & Autoencoder Latent & Additional Module & \# of params & Accuracy  \\
& Dimension & Layers &  & Dimension &  &  &  (\%) \\
    \midrule
A1 & 256 & 2 & [32,64,32] & 128 & – & 669,368 & 92.25\\
A2 & 256 & 1 & [32,64,32] & 128 & – & 537,784 & 93.09\\
A3 & 256 & 1 & [48,64,48] & 128 & – & 557,880 & 95.30\\
A4 & 256 & 1 & [48,64,48] & 64 & – & 508,664 & 95.35 \\
A5 & 256 & 1 & [48,64,48] & 128 & Cross-modal Attention & 681,528 &  90.46\\
A6 & 128 & 2 & [32,64,32] & 128 & – & 447,544 & 88.94 \\
A7 & 128 & 1 & [32,64,32] & 128 & – & 414,520 &  94.88\\
A8 & 128 & 1 & [48,64,48] & 128 & – & 432,568 & \textbf{95.45}\\
A9 & 128 & 1 & [32,64,32] & 64 & – & 365,304 & 91.25 \\
A10 & 128 & 1 & [48,64,48] & 64 & – & 383,352 & 94.17\\
A11 & 128 & 1 & [48,64,48] & 128 & Cross-modal Attention & 539,832 & 86.24 \\
     \bottomrule
    \end{NiceTabular}

\end{table*}

\begin{table*}[htbp]
    \caption{Ablation study based on modality}
    \label{Modality ablation study}
    \centering
    \begin{NiceTabular}{c c c c c c c c c c | c c c c | c}
    \toprule
    \multirow{2}{*}{ID} & \multicolumn{5}{c}{EEG} & \multirow{2}{*}{HRV} & \multirow{2}{*}{EDA} & \multirow{2}{*}{Temp} & \multirow{2}{*}{Personality} & \# of params & FLOPS & Latency & Size & Accuracy  \\
\cmidrule{2-6}
    & Raw & PSD & DE & Stats. & Asym. &&&&& (K) & (M) & (ms/samp.) & (MB) & (\%)\\
    \midrule
    A1 & \ding{51} & \ding{51} & \ding{51} & \ding{51} & \ding{51} & \ding{51} & \ding{51} & \ding{51} & \ding{51} & 432.6 & 6.636 & 0.1716 & 1.6501 & 95.45 \\ 
    A2 &\ding{51} & \ding{51} & \ding{51} & \ding{51} & \ding{51} & \ding{53} & \ding{53} & \ding{53} & \ding{53} & 387.7 & 6.591 &  0.1573 & 1.4788 & 85.16 \\ 
    A3 & \ding{51} & \ding{51} & \ding{51} & \ding{51} & \ding{51} & \ding{53} & \ding{53} & \ding{53} & \ding{51} & 396.6 & 6.601 & 0.1534 & 1.5129 & 96.04 \\ 
    A4 & \ding{53} & \ding{53} & \ding{53} & \ding{53} & \ding{53} & \ding{51} & \ding{51} & \ding{51} & \ding{51} & 178.7 & 0.178 & 0.0165 & 0.6815 & 29.05 \\
    A5 & \ding{51} & \ding{53} & \ding{53} & \ding{53} & \ding{53} & \ding{53} & \ding{53} & \ding{53} & \ding{53} & 274.6 & 6.479 & 0.1457 & 1.0474 & 86.89 \\
    A6 & \ding{51} & \ding{53} & \ding{53} & \ding{53} & \ding{53} & \ding{53} & \ding{53} & \ding{53} & \ding{51} & 274.6 & 6.479 & 0.1454 & 1.0474 & 96.35 \\
    A7 & \ding{53} & \ding{51} & \ding{51} & \ding{51} & \ding{51} & \ding{51} & \ding{51} & \ding{51} & \ding{51} & 291.7 & 0.291 & 0.0239 & 1.113 & 74.44 \\
    A8 & \ding{51} & \ding{51} & \ding{53} & \ding{53} & \ding{53} & \ding{53} & \ding{53} & \ding{53} & \ding{53} & 297.6 & 6.502 & 0.1521 & 1.1352 & 82.48 \\
    A9 & \ding{51} & \ding{51} & \ding{53} & \ding{53} & \ding{53} & \ding{53} & \ding{53} & \ding{53} & \ding{51} & 306.5 & 6.511 & 0.1525 & 1.169 & 95.90 \\
    A10 & \ding{51} & \ding{53} & \ding{51} & \ding{53} & \ding{53} & \ding{53} & \ding{53} & \ding{53} & \ding{53} & 297.6 & 6.502 & 0.1519 & 1.1352 & 89.51 \\
    A11 & \ding{51} & \ding{53} & \ding{51} & \ding{53} & \ding{53} & \ding{53} & \ding{53} & \ding{53} & \ding{51} & 306.5 & 6.511 & 0.1508 & 1.1693 & 96.76 \\
    A12 & \ding{51} & \ding{53} & \ding{53} & \ding{51} & \ding{51} & \ding{53} & \ding{53} & \ding{53} & \ding{53} & 341.5 & 6.546 & 0.1519 & 1.3031 & 67.32 \\
    A13 & \ding{51} & \ding{53} & \ding{53} & \ding{51} & \ding{51} & \ding{53} & \ding{53} & \ding{53} & \ding{51} & 350.5 & 6.555 & 1.3371 & 0.1535 & 94.32 \\
    \bottomrule
    \end{NiceTabular}
\end{table*} 

We have employed AdamW optimization~\cite{loshchilov2017decoupled} with a decoupled weight decay of $0.01$ to ensure stable convergence. Training is initialized by a 15-epoch linear warmup followed by Cosine Annealing~\cite{loshchilov2017sgdrstochasticgradientdescent}. To mitigate the inherent noise in self-reported emotion labels and the imbalance of the dataset, the loss function integrates Label Smoothing (0.1) and Balanced Class Weights, alongside an Adaptive Autoencoder Weight ($\lambda_{AE}$) that decays linearly to shift the model's focus from unsupervised representation learning to supervised classification. Finally, training is guarded by gradient clipping (1.0) and Early Stopping, with a patience of 25 based on the Macro F1-score, to prevent overfitting. Note that all model configurations reported in Table~\ref{tab:hyperparameter}, as well as in all the subsequent tables, were trained for 200 epochs.

\subsection{Ablation Study}

The ablation study is presented in two stages: the first stage includes an architectural ablation study presented in Table~\ref{Architecture ablation}, and the second stage is based on modalities presented in Table~\ref{Modality ablation study}. The architectural ablation study is based on all the modalities included. Note that all models were trained using identical optimization settings to ensure fair comparison. The architectural ablation study results show that the inclusion of cross-modal attention after feature concatenation negatively impacted the accuracy. The architecture showed sensitivity to the width of CNN layers compared to the depth of liquid layers, with wider CNN layers leading to better accuracy. The best configuration, A8, leading to an accuracy of 95.45\%, consists of CNN filters with size [48, 64, 48] followed by one liquid layer with the hidden dimension of 128, with an autoencoder hidden dimension to be 128, as well as cross-modal attention. 

Based on our findings from the architectural ablation study, we performed a detailed ablation study of different modalities of data. The latency of each case is based on the model's performance on the Nvidia Tesla dual T4 GPU. The ablation study presented in Table~\ref{Modality ablation study} provides insight into the contribution of various physiological and psychometric modalities to the overall performance of our proposed architecture. The full model configuration achieves an accuracy of 95.45\%, utilizing a total parameter count of 432.6K. However, the highest recorded accuracy of 96.76\% is achieved by combining raw EEG, DE, and personality traits. This suggests that high-level spectral features like DE, when coupled with static personality profiles, provide a more discriminative feature space than the inclusion of all peripheral sensors combined. 

A critical finding is the significant impact of the personality modality on accuracy. Adding Personality traits to raw EEG increases accuracy from 86.89\% to 96.35\% with negligible impact on latency (~0.145 ms). Besides, an increase in accuracy after the inclusion of personality traits is also seen when PSD, DE, statistical features, and asymmetric features are considered. This highlights that emotional responses are highly dependent on individual traits, and the model successfully leverages this static context to resolve ambiguities in dynamic EEG signals. 

\begin{figure*}[htbp]
     \centering
     \begin{subfigure}[b]{0.45\textwidth}
         \centering
         \includegraphics[width=\textwidth]{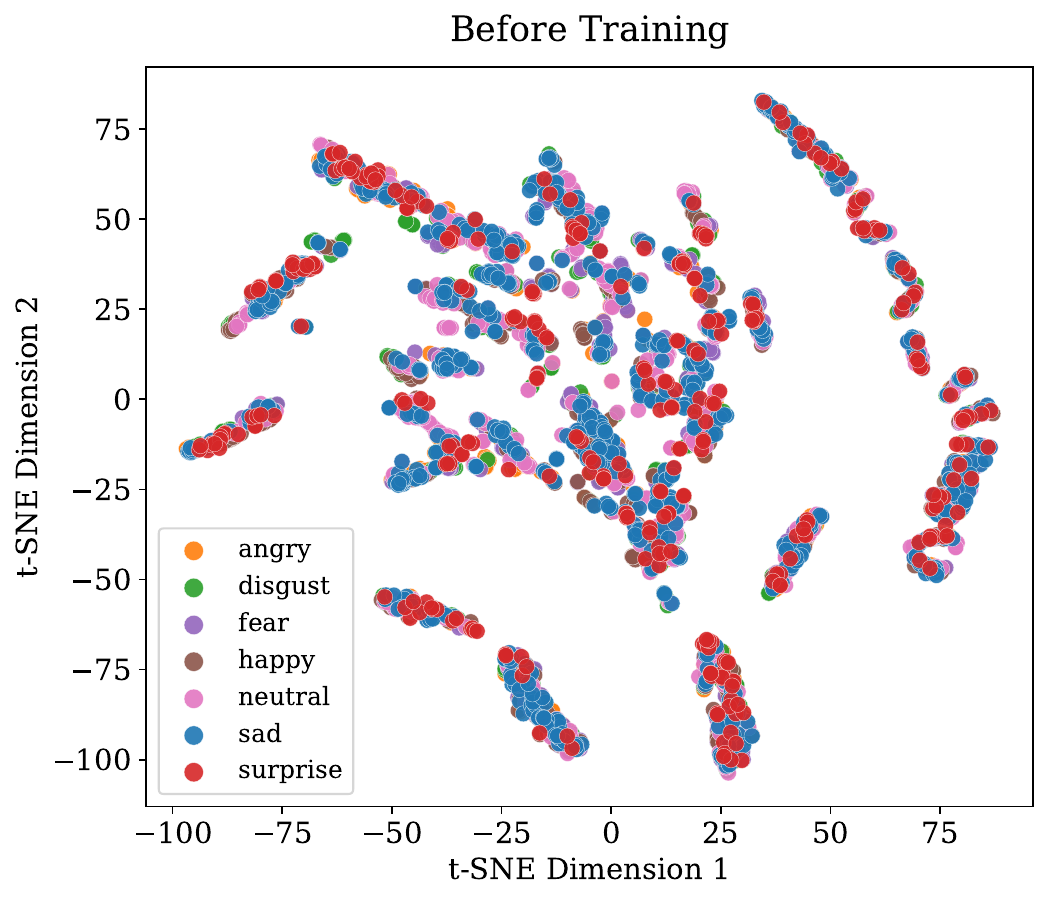}
         \caption{t-SNE plot of 4000 samples before training}
         \label{fig:image1}
     \end{subfigure}
     \hfill
     \begin{subfigure}[b]{0.45\textwidth}
         \centering
         \includegraphics[width=\textwidth]{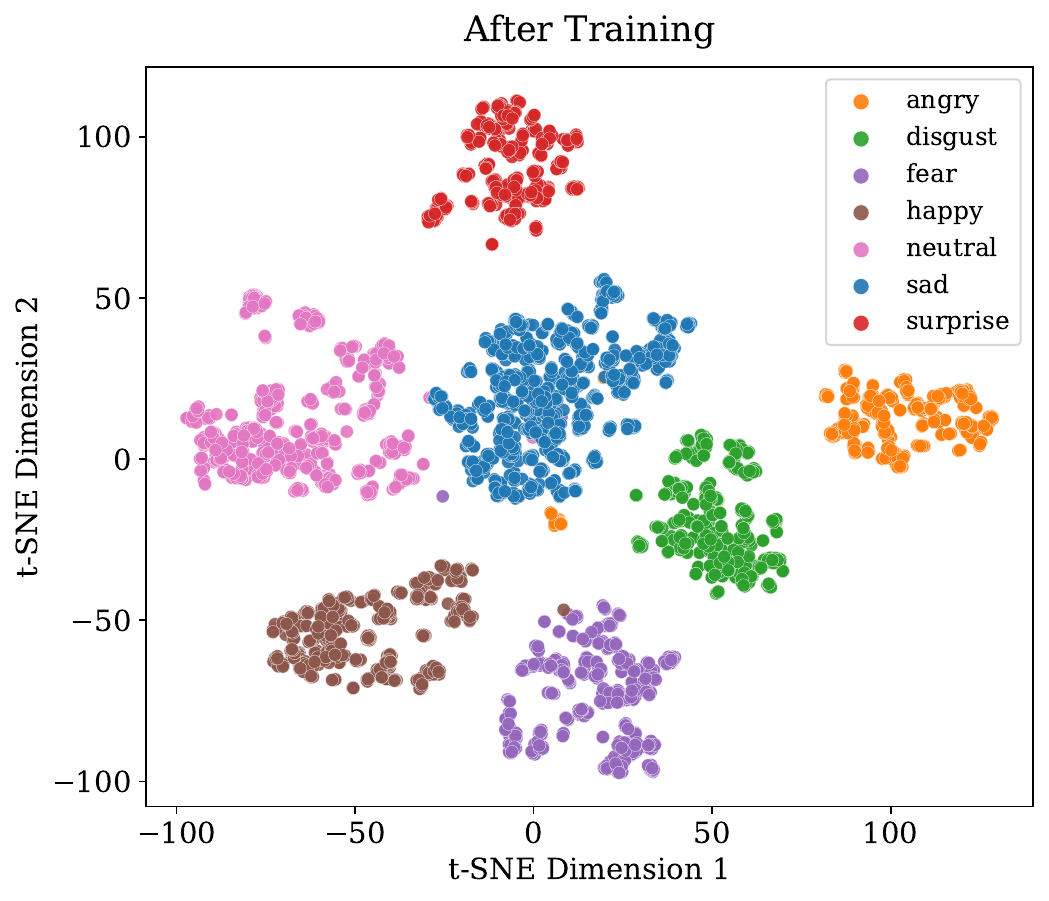}
         \caption{t-SNE plot of 4000 samples after training}
         \label{fig:image2}
     \end{subfigure}
     \caption{t-SNE plot before and after training}
     \label{fig:tsne}
\end{figure*} 

\subsection{Interpretability of Our Proposed Architecture}

\subsubsection{t-SNE Plot}

To visualize the high-dimensional latent features learned by the model, we employ t-Distributed Stochastic Neighbor Embedding (t-SNE)~\cite{van2008visualizing}, which projects data into a two-dimensional space while preserving local neighborhood relationships. This allows an intuitive understanding of class structure and separability in the learned feature space. In the seven-class emotion setting, the t-SNE plots in Figure~\ref{fig:tsne} show a clear difference before and after training. Prior to training, emotion samples are highly mixed, indicating poor class discrimination. After training, the embeddings form compact and well-separated clusters, demonstrating that the proposed multimodal liquid–attention framework learns emotion-specific representations effectively.

\begin{table}[htbp]
    \centering
    \caption{Cluster Separability Metrics (t-SNE space) before and after training}
    \label{tab:cluster metrics}
    \begin{tabular}{l r r}
\toprule
Metrics & Before Training & After Training \\
\midrule
Calinski-Harabasz ↑ & 0.618 & 431.362 \\
Davies-Bouldin ↓ & 69.216 & 1.745\\
Inter-Centroid (Euclidean) ↑ & 0.111 & 0.904\\
Inter-Centroid (Mahalanobis) ↑ & 1.242 & 213.122 \\
\bottomrule
\end{tabular}
\end{table}

Here four standard metrics are used to quantitatively evaluate cluster quality: Calinski–Harabasz (CH) Index~\cite{calinski1974dendrite}, Davies–Bouldin (DB) Index~\cite{4766909}, Inter-Centroid Euclidean Distance, and Inter-Centroid Mahalanobis Distance~\cite{mahalanobis2018generalized}. As shown in Table~\ref{tab:cluster metrics}, training leads to substantial improvement in all metrics. The CH index increases from 0.618 to 431.362, while the DB index drops from 69.216 to 1.745, indicating significantly improved cluster compactness and reduced overlap. Additionally, both Euclidean (0.111 → 0.904) and Mahalanobis (1.242 → 213.122) distances between emotion centroids increase, confirming stronger separation in the latent space. These results validate that the proposed training strategy produces more discriminative emotion representations.

\subsubsection{Temporal Attention Analysis}

\begin{figure*}[htbp]
     \centering
     \begin{subfigure}[b]{0.44\textwidth}
         \centering
         \includegraphics[width=\textwidth]{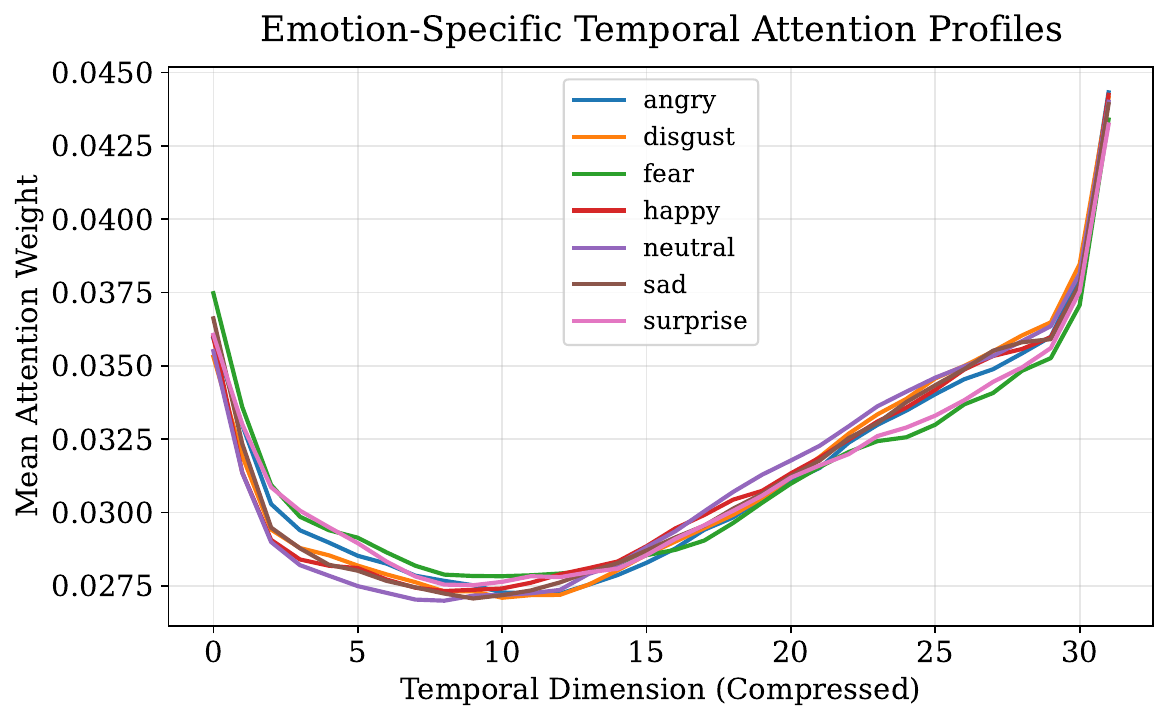}
         \caption{Emotion-wise Attention Curves}
         \label{fig:emotionwise attention curve}
     \end{subfigure}
     \hfill 
     \begin{subfigure}[b]{0.54\textwidth}
         \centering
         \includegraphics[width=\textwidth]{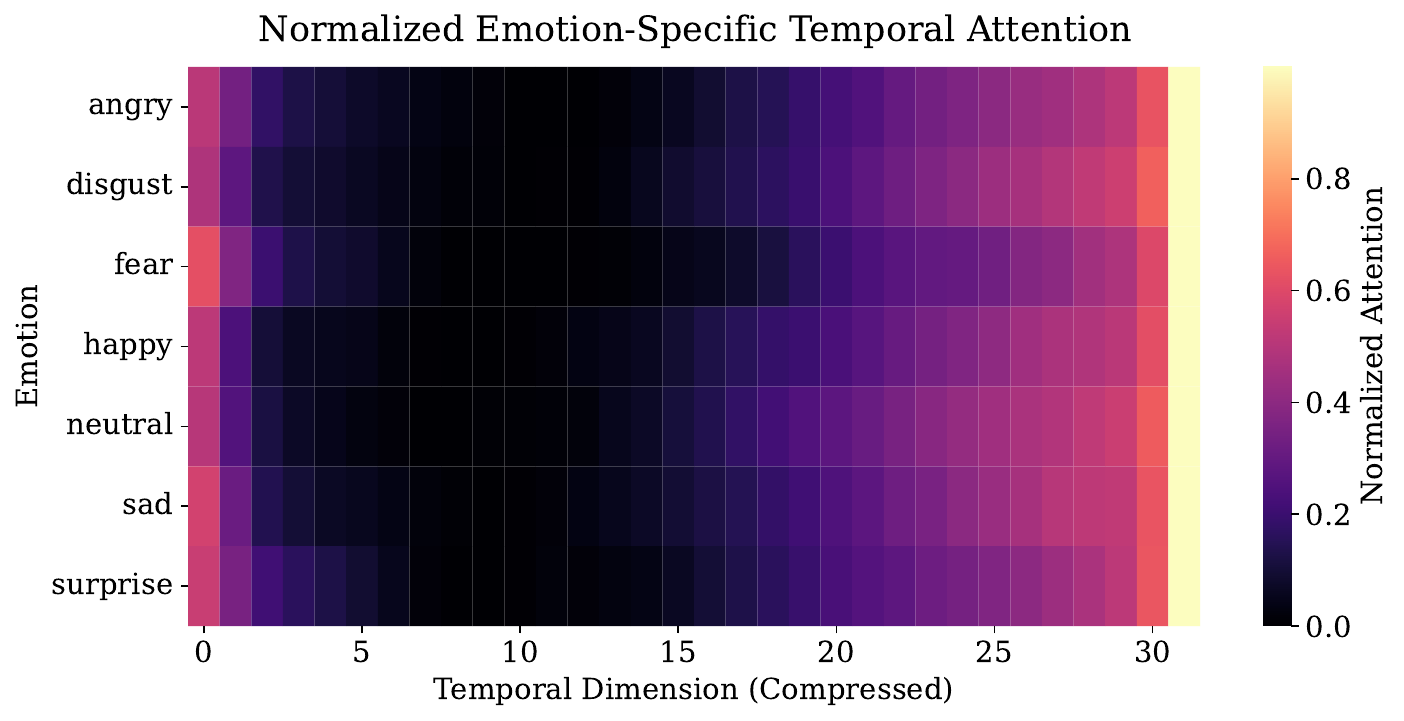}
         \caption{Normalized Emotion-wise Attention Heatmap}
         \label{fig:normalised attention heatmap}
     \end{subfigure}
     \caption{Temporal Attention Analysis of the Liquid Neural Network}
     \label{fig:temporal attention}
\end{figure*}

One of the major contributions of our work is to introduce an LNN with learnable time constants for emotion detection. To analyze how our model focuses on features in the temporal domain, we examine the mean attention weights for individual emotions at different time steps. Figure~\ref{fig:temporal attention} presents an analysis of the temporal attention behavior learned by our architecture. Figure~\ref{fig:emotionwise attention curve} illustrates the emotion-specific temporal attention profiles, where each curve represents the mean attention weight aggregated over all samples belonging to a specific emotional class. The horizontal axis represents the temporal sequence length after CNN-based feature extraction. Specifically, raw EEG signals (14 channels × 256 time points) undergo successive 1D convolution and max-pooling operations through three CNN layers, reducing the temporal dimension from 256 to 32 time steps. The LNN then processes this compressed sequence, and attention weights are computed at each of the 32 temporal positions. While this compressed axis does not preserve absolute temporal resolution in seconds, it maintains the relative temporal ordering of neural dynamics within each trial, with earlier positions corresponding to earlier portions of the EEG signal.

An interesting observation can be found across all emotional categories, which is a consistent U-shaped attention pattern. This pattern likely suggests that our model focuses on two distinct temporal components of EEG activity.
\begin{itemize}
    \item Early transient responses: These likely correspond to initial event-related potentials (ERPs) following emotional stimulus presentation, encoding rapid sensory and evaluative processing ~\cite{OLOFSSON2008247}.
    \item Late sustained responses: These likely reflect a sustained emotional processing, working memory maintenance, and regulatory processes~\cite{Hajcak12022010}
\end{itemize}

Figure~\ref{fig:normalised attention heatmap} presents the normalized emotion-wise attention heatmap, which reveals fine-grained temporal differences between emotional categories. Emotions like fear, anger, sadness, and surprise show strong attention, whereas happiness, disgust, and neutral states shointermediate-late-stagege attention profiles. These distinct temporal signatures indicate that different emotional states are characterized by unique temporal dynamics in EEG activity, which our architecture effectively captures through its learned attention mechanism. However, given that our 2-second EEG segments are not stimulus-locked, this pattern can also likely be caused by architectural biases from CNN max-pooling operations rather than stimulus-driven temporal dynamics.

\subsection{Statistical Analysis of Temporal Dynamics}

\begin{figure*}[htbp]
    \centering
    \begin{subfigure}[b]{0.48\textwidth}
        \centering
        \includegraphics[width=\linewidth]{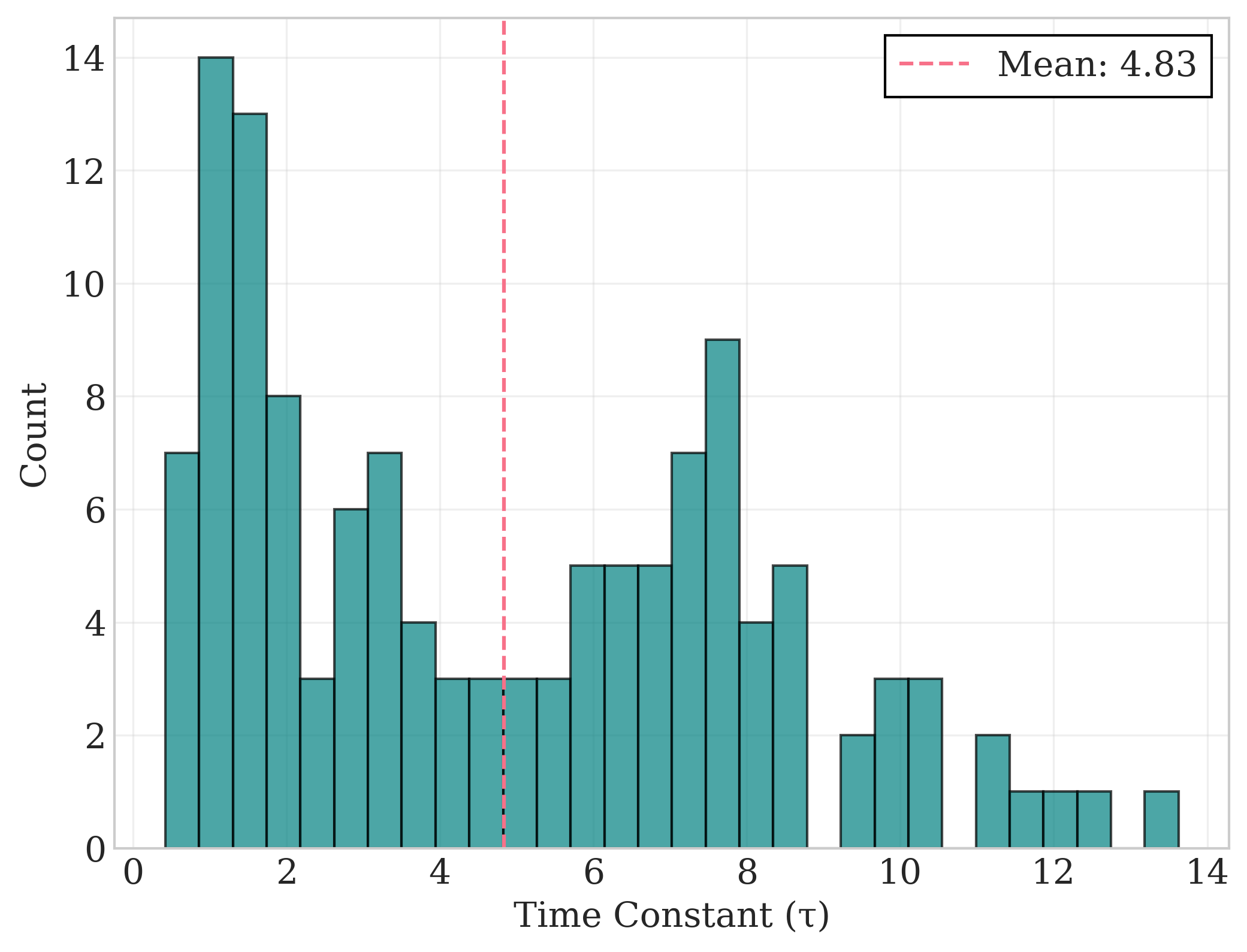}
        \caption{}
        \label{fig:tau_distribution}
    \end{subfigure}
    \hfill
    \begin{subfigure}[b]{0.48\textwidth}
        \centering
        \includegraphics[width=\linewidth]{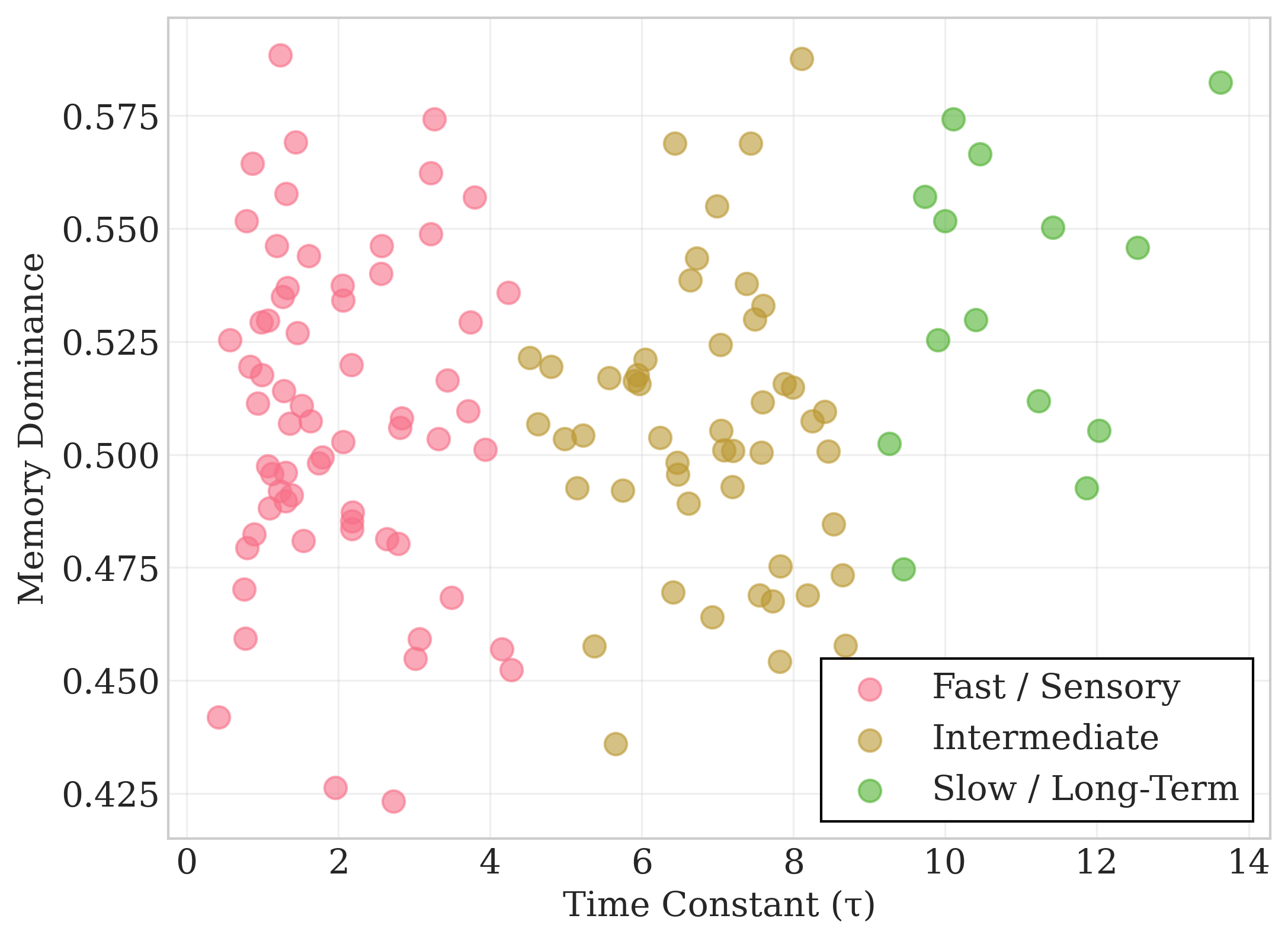}
        \caption{}
        \label{fig:memory_dominance_time_constant}
    \end{subfigure}
    \caption{Liquid Neural Network dynamics: (a) Time constant ($\tau$) distribution, (b) Memory dominance for different time constants across three neuron types}
    \label{fig:liquid_analysis}
\end{figure*}

In this section, we validate the adaptive temporal dynamics of our proposed network through various statistical analyses. Figure~\ref{fig:tau_distribution} shows the distribution of learned time constants across 128 neurons in our LNN. To check whether the time constants vary randomly across a mean, we conducted a Shapiro-Wilk test~\cite{10.1093/biomet/52.3-4.591} on log-transformed $\tau$ values. The result indicates significant deviation from log-normality ($W = 0.93, p < 0.0001$), suggesting the distribution reflects distinct functional neuron populations rather than a single continuous process. 
Figure~\ref{fig:memory_dominance_time_constant} shows the relationship between learned time constants and memory dominance across 128 individual neurons in our network. By using K-means clustering~\cite{8ddb7f85-9a8c-3829-b04e-0476a67eb0fd} (k=3), we identified three categories which we denote as Fast/sensory neurons for rapid processing, intermediate neurons for feature integration, and slow/long-term neurons for temporal context maintenance. We note that this heterogeneous temporal hierarchy emerged naturally through backpropagation without architectural constraints. This demonstrates the network has naturally evolved into distinct groups, where different neurons specialize in processing information at different timescales. 

\begin{table}[htbp]
    \centering
    \caption{Neuron role identification through K-means clustering}
    \label{tab:neuron role}
    \begin{tabular}{l c c c}
    \toprule
    \textbf{Role} & \textbf{Count} & \textbf{Avg. $\tau$} & \textbf{Avg. Mem. Dom.}  \\
    \midrule
    Fast / Sensory & 65 & 2.0080 & 0.5084 \\        
    Intermediate & 49 & 6.8596 & 0.5049 \\      
    Slow / Long-Term & 14 & 10.8579 & 0.5336 \\
    \bottomrule
    \end{tabular}

\end{table}

Table~\ref{tab:neuron role} shows the number of neurons for each category, their average time constant,t, and memory dominance. Memory dominance quantifies the relative strength of recurrent versus feedforward processing in neural networks~\cite{pmlr-v28-pascanu13}. It is defined as,

\begin{equation}
  \text{MD} = \frac{||W_{rec}||_F}{||W_{rec}||_F + ||W_{in}||_F}
\end{equation}

where $W_{rec}$ and $W_{in}$ represent recurrent and input weight matrices, and $||\cdot||_F$ denotes the Frobenius norm. Values above 0.5 indicatea preference for temporal integration over instantaneous processing. For our network, we observe that fast/sensory neurons exhibit average memory dominance at 0.5084, suggesting balanced feedforward-recurrent processing for rapid responses. Intermediate neurons show similar memory dominance, indicating consistent integration strategies across moderate timescales. Slow/long-term neurons display slightly elevated memory dominance, reflecting modest increased reliance on recurrent processing for maintaining temporal context. Finally, Spearman correlation analysis~\cite{Spearman_1904} reveals no significant relationship between $\tau$ and memory dominance ($\rho = 0.063, p = 0.48$). This indicates that these two network properties emerge independently during training. The time constants determine integration timescales while memory dominance reflects the balance between input and recurrent processing. This proves that the LNN can independently tune how long a neuron retains information (via $\tau$) and how much it prioritizes internal state over external stimuli (via memory dominance).

\begin{figure}[htbp]
    \centering
    \includegraphics[width=1\linewidth]{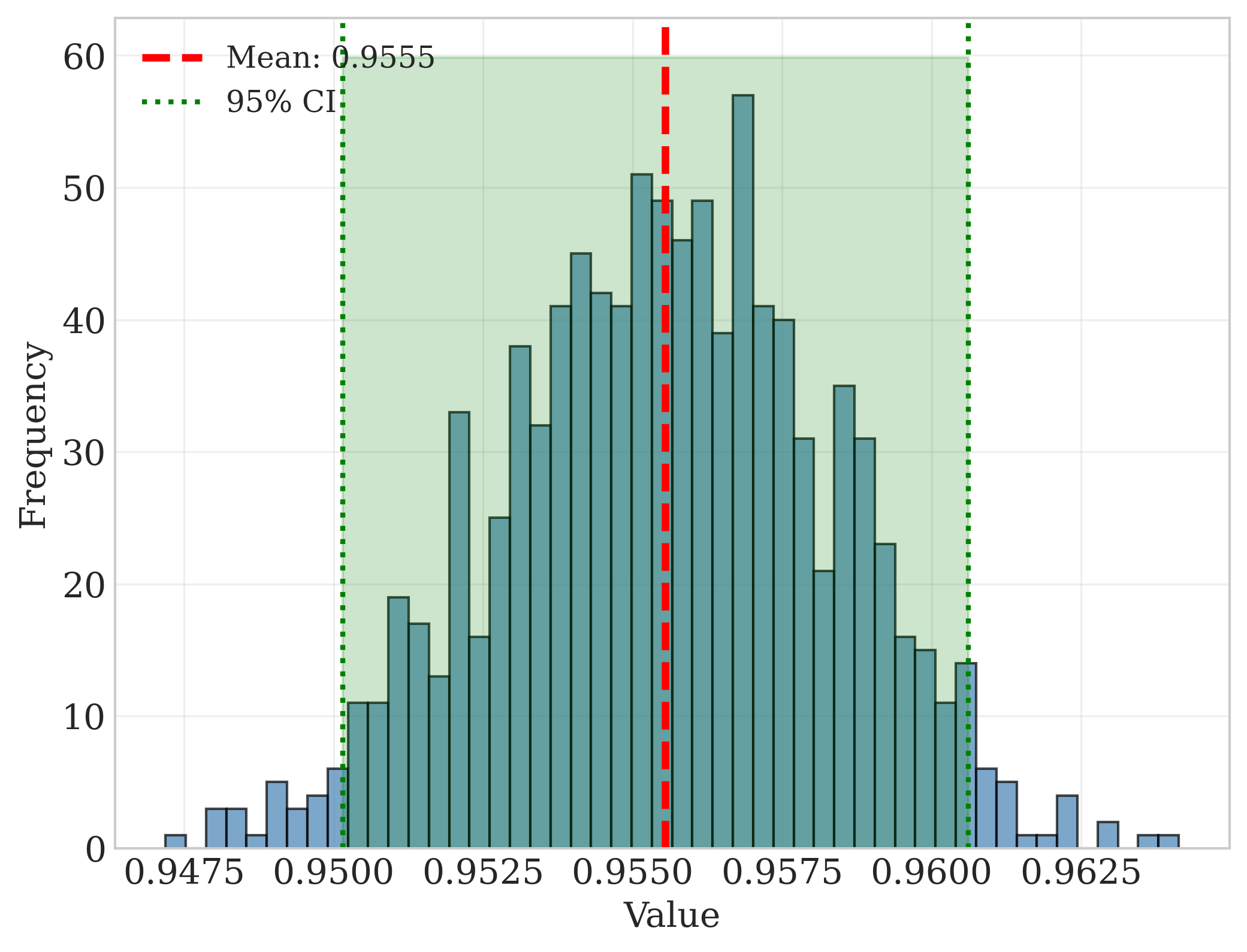}
    \caption{Bootstrap distribution of accuracy over 1,000 iterations with 95\% CI}
    \label{fig:bootstrap_accuracy}
\end{figure}

Figure~\ref{fig:bootstrap_accuracy} displays the bootstrap distribution of classification accuracy over 1,000 resampling iterations. The narrow 95\% confidence interval [95.01\%, 96.06\%] demonstrates high stability, indicating that performance estimates are robust to sampling variability~\cite{carpenter2000bootstrap}. The approximately Gaussian distribution of bootstrap samples confirms the statistical validity of the reported metrics.

Beyond classification accuracy, the reliability of predicted probabilities is also important for real-world deployment. Table~\ref{tab:calibration} summarizes the model calibration metrics. The Expected Calibration Error (ECE) of 0.0812 signifies medium calibration, where predicted confidence levels reasonably approximate true correctness probabilities~\cite{guo2017calibration}. The Brier score of 0.0102 is substantially below the 0.25 threshold for random guessing. This confirms excellent probability estimation~\cite{glenn1950verification} and the mean prediction confidence of 0.882 indicates decisive classifications. However, the Maximum Calibration Error (MCE) of $0.2347$ highlights a reliability gap in the model's least calibrated confidence bin. The MCE value represents a worst-case deviation where predicted confidence and empirical accuracy diverge by $23.4\%$. This suggests that despite an excellent Brier Score, the model exhibits localized overconfidence. This behavior is likely to occur at the decision boundaries of minority functional classes. To mitigate this peak error, future work utilizing temperature scaling~\cite{guo2017calibration} or Platt scaling~\cite{platt1999probabilistic} could be employed, ensuring that the network's predictive uncertainty is robust enough for clinical integration.

\begin{table}[htbp]
    \centering
    \caption{Model calibration metrics}
    \label{tab:calibration}
    \begin{tabular}{l l}
        \toprule
        \textbf{Metric} & \textbf{Value} \\
        \midrule
        Expected Calibration Error (ECE) & 0.0812\\
        Mean Calibration Error (MCE) & 0.2347 \\
        Brier Score & 0.0102 \\
        Mean Confidence & 0.882 \\
        \bottomrule
    \end{tabular}
\end{table}

\subsection{Comparison with Existing Works}

We have compared our architecture in two separate aspects. First, we have compared our LNN for raw EEG with other existing architectures used for temporal modeling.  

\begin{table}[htbp]
    \centering
    \caption{Comparative analysis of temporal modeling mechanisms under identical multimodal feature extraction and fusion settings.}
    \label{tab:Temporal Modelling Comparison}
    \begin{tabular}{l r r r}
    \toprule
Temporal Architecture & Test Acc & Test F1 & \# of params \\
\midrule
LNN & 0.9545 & 0.9527 & \textbf{432,568}\\
LSTM~\cite{hochreiter1997long} & \textbf{0.9610} & \textbf{0.9564} & 500,920\\
GRU~\cite{cho2014learning} & 0.9502 & 0.9481 & 478,136\\
Temporal CNN~\cite{bai2018empirical} & 0.9211 & 0.9191 & 445,112\\
\bottomrule
\end{tabular}
\end{table}

\begin{table*}[htbp]
    \centering
    \caption{Comparative analysis of our work with prior studies on the PhyMER dataset for subject-dependent performance}
    \label{tab:comparison with works}
    \begin{tabular}{l l l l l l}
    \toprule
    Study/Reference & \# of classes & \# of params (M) & Accuracy (\%) & F1-score (\%) & MCC (\%) \\
    \midrule
    Wu et al.~\cite{wu2025phy} & 3 & - & 84.2 & - & -\\
    Jiang et al.~\cite{jiang2023multimodal} & 4 & \textbf{0.074} & 67.98 & 64.95 & -\\
    Jia et al.~\cite{jia2021hetemotionnet} & 4 & 0.261 & 68.13 & 63.68 & -\\
    Xu et al.~\cite{xu2025emotion} & 4 & 0.388 & 72.30 & 67.69 & -\\
    Kim et al.~\cite{kim2025mifu} & 7 & 0.54 & 73.06 & - & -\\
    Pant et al.~\cite{pant2023phymer} & 7 & - & - & 76.73 & 72.55\\
    Our proposed architecture & 7 & 0.432 & \textbf{95.45} & \textbf{93.71} & \textbf{94.37} \\
    \bottomrule
    \end{tabular}
\smallskip
\small
\\
\textit{Note:} '–' indicates that the value was not reported in the original study.
\end{table*}

In Table~\ref{tab:Temporal Modelling Comparison}, we have presented the accuracy score for four different types of temporal modeling architectures while keeping everything else equivalent in all four architectures. While LSTM achieves marginally higher accuracy, it requires more parameters and lacks post-deployment adaptability. Architectures like GRU or Temporal CNN require fewer parameters than the LSTM, but also show a drop in accuracy. While TCNs and transformers capture multi-scale dependencies through architectural depth or dilation, LNNs explicitly model without increasing depth or parameter count. To this end, the proposed LNN provides a favorable trade-off between performance, parameter efficiency, and adaptability. Unlike fixed recurrent architectures, LNNs allow continuous temporal adaptation through dynamic state evolution without explicit weight updates. This makes the LNN architecture suitable for edge device deployment with preserved accuracy.
Finally, we compare our proposed architecture with existing works on the PhyMER dataset and present the results in Table~\ref{tab:comparison with works}.

Our results show that the LNN architecture is uniquely suited for subject-dependent deployment on resource-constrained edge devices. In such applications, the hardware is often dedicated to a single user. This makes subject-independent generalization less important than localized precision and low power consumption. By focusing on subject-dependent dynamics, the LNN significantly outperforms larger models in the literature. This suggests that for personalized devices, the liquid state's ability to adapt to an individual’s specific neural signatures provides a superior performance-to-size ratio compared to traditional fixed-parameter architectures.

\section{Discussion}

To the best of our knowledge, this is the first multimodal EEG-based emotion recognition framework that integrates LNNs with learnable time constants, attention-guided fusion, and peripheral physiological signals. In this section, we discuss our design decision, analyze the implications of our results and deployment issues, acknowledge our limitations, and outline future research directions. 

\textbf{Design Decisions and Architectural Novelty:} The implementation of the CNN-LNN-Attention pipeline represents a significant shift from static temporal modeling to adaptive, bio-inspired computation. Traditional RNNs, including LSTMs and GRUs, have dominated time-series modeling in emotion detection due to their ability to capture sequential dependencies through discrete-time updates. However, these models often struggle with long-range dependencies, vanishing gradients, and sensitivity to noise in multimodal data streams, which are prevalent in emotion recognition tasks. LNNs, introduced as LTC networks, address these limitations by modeling neural dynamics as continuous-time ordinary differential equations, enabling adaptive temporal scales~\cite{zong2025accuracy}. Our preprocessing strategy, which incorporates ICA-based artifact rejection and 2s windowing, was specifically engineered to ensure that the LNN receives a high-SNR (Signal-to-Noise Ratio) input required for reliable spectral analysis. Unlike traditional approaches that rely solely on raw signals or manual features, our framework treats different features as complementary. While PSD and DE provide artifacts of spectral power, the LNN layers capture the non-stationary transitions inherent in EEG dynamics. The use of specialized MLP subnetworks for peripheral signals (BVP, EDA, TEMP) serves as an efficient dimensionality reduction step, ensuring that the final autoencoder-based fusion module can learn complex, cross-modal correlations without being overwhelmed by the high dimensionality of the raw input space. Results in Table~\ref{Modality ablation study} show that the architecture maintains a sub-millisecond latency (<0.18 ms/sample) across nearly all configurations. The most complex model with all modalities requires 6.636M FLOPs, yet achieves a high-speed inference capability that is well-suited for real-time Brain-Computer Interface applications. The minimal size (~1.65 MB) further underscores the potential for deployment on edge devices with limited memory.

\textbf{Implications of Results and Subject-Dependence:} The observed subject-dependent accuracy of 95.45\% underscores the LNN's ability to tune its internal time constants to the specific neurophysiological signatures of an individual. This is particularly relevant for personalized nanoelectronics and wearable EEG headbands, where the device must adapt to the unique neural dialect of the user. By utilizing an autoencoder for latent representation learning, we ensure that the features being classified are statistically salient and stripped of modality-specific noise. However, the high parameter count of the autoencoder ($53\%$ of total params) suggests that while performance is maximized, there is a remaining opportunity to explore more sparse fusion techniques.

\textbf{Limitations and Future Directions:} A primary limitation of the current study is the exclusion of subject-independent analysis. While LNNs are inherently suited for the temporal nuances of subject-dependent data, the cross-subject variability in EEG remains a challenge for global generalization. The inclusion of such subject-specific features into a subject-independent classification environment can be a promising direction for research. Additionally, replacing the autoencoder with a more lightweight attention-based fusion mechanism, such as bilinear pooling or cross-attention, could reduce the memory footprint, facilitating the transition from high-performance workstations to real-time, on-chip emotion monitoring systems.

\section{Conclusion}
In this work, we proposed a multimodal emotion recognition architecture that utilizes the adaptive temporal modeling capabilities of LNNs, combined with attention mechanisms and autoencoder-based feature fusion. By utilizing raw EEG signals, handcrafted EEG features, peripheral physiological features, and personality traits, the proposed model captures both short-term and long-range affective dynamics. The incorporation of learnable time constants enables the network to adapt to the non-stationary nature of emotional responses.

Extensive experimental evaluations show that the proposed architecture achieves strong classification performance across seven emotional categories. Architecture and hyperparameter ablation studies further confirm the importance of liquid dynamics, CNN front-end design, and latent fusion dimensionality. Visualization of latent representations using t-SNE, along with quantitative compactness metrics, reveals significantly improved class separability after training. Statistical analysis of the learned network dynamics shows that neurons self-organized into functionally distinct temporal processing units. Further memory dominance analysis indicates that the network prioritizes modeling temporal dependencies within observation windows over instantaneous input transformations. In addition, temporal attention analysis highlights consistent global trends and emotion-specific temporal emphasis patterns, reinforcing the interpretability of the learned representations.

Overall, this work shows that attention-augmented LNNs can provide an effective and interpretable approach for personalized multimodal emotion recognition. Future work may explore subject-adaptive liquid dynamics, cross-dataset generalization, and lightweight architectures to further enhance robustness in real-world affective computing applications.

\section*{Declaration of competing Interest}
The authors declare that they have no known competing financial interests or personal relationships that could have appeared to influence the work reported in this paper.

\ifCLASSOPTIONcaptionsoff
\newpage
\fi

\bibliographystyle{ieeetr}
\bibliography{reference}

@article{pant2023phymer,
  title={PhyMER: physiological dataset for multimodal emotion recognition with personality as a context},
  author={Pant, Sudarshan and Yang, Hyung-Jeong and Lim, Eunchae and Kim, Soo-Hyung and Yoo, Seok-Bong},
  journal={IEEE Access},
  volume={11},
  pages={107638--107656},
  year={2023},
  publisher={IEEE}
}

@article{coan2004frontal,
  title={Frontal EEG asymmetry as a moderator and mediator of emotion},
  author={Coan, James A and Allen, John JB},
  journal={Biological psychology},
  volume={67},
  number={1-2},
  pages={7--50},
  year={2004},
  publisher={Elsevier}
}

@article{zhao2017emotion,
  title={Emotion analysis for personality inference from EEG signals},
  author={Zhao, Guozhen and Ge, Yan and Shen, Biying and Wei, Xingjie and Wang, Hao},
  journal={IEEE transactions on affective computing},
  volume={9},
  number={3},
  pages={362--371},
  year={2017},
  publisher={IEEE}
}

@article{welch2003use,
  title={The use of fast Fourier transform for the estimation of power spectra: A method based on time averaging over short, modified periodograms},
  author={Welch, Peter},
  journal={IEEE Transactions on audio and electroacoustics},
  volume={15},
  number={2},
  pages={70--73},
  year={2003},
  publisher={IEEE}
}

@article{winkler2011automatic,
  title={Automatic classification of artifactual ICA-components for artifact removal in EEG signals},
  author={Winkler, Irene and Haufe, Stefan and Tangermann, Michael},
  journal={Behavioral and brain functions},
  volume={7},
  number={1},
  pages={30},
  year={2011},
  publisher={Springer}
}

@article{albera2012ica,
  title={ICA-based EEG denoising: a comparative analysis of fifteen methods},
  author={Albera, Laurent and Kachenoura, Amar and Comon, Pierre and Karfoul, Ahmad and Wendling, Fabrice and Senhadji, Lotfi and Merlet, Isabelle},
  journal={Bulletin of the Polish Academy of Sciences: Technical Sciences},
  volume={60},
  number={3 Special issue on Data Mining in Bioengineering},
  pages={407--418},
  year={2012}
}

@article{uriguen2015eeg,
  title={EEG artifact removal—state-of-the-art and guidelines},
  author={Urig{\"u}en, Jose Antonio and Garcia-Zapirain, Bego{\~n}a},
  journal={Journal of neural engineering},
  volume={12},
  number={3},
  pages={031001},
  year={2015},
  publisher={IOP Publishing}
}

@article{delorme2007enhanced,
  title={Enhanced detection of artifacts in EEG data using higher-order statistics and independent component analysis},
  author={Delorme, Arnaud and Sejnowski, Terrence and Makeig, Scott},
  journal={Neuroimage},
  volume={34},
  number={4},
  pages={1443--1449},
  year={2007},
  publisher={Elsevier}
}

@inproceedings{garcia2017emotion,
  title={Emotion detection: a technology review},
  author={Garcia-Garcia, Jose Maria and Penichet, Victor MR and Lozano, Maria D},
  booktitle={Proceedings of the XVIII international conference on human computer interaction},
  pages={1--8},
  year={2017}
}

@article{russell1980circumplex,
  title={A circumplex model of affect.},
  author={Russell, James A},
  journal={Journal of personality and social psychology},
  volume={39},
  number={6},
  pages={1161},
  year={1980},
  publisher={American Psychological Association}
}

@book{picard2000affective,
  title={Affective computing},
  author={Picard, Rosalind W},
  year={2000},
  publisher={MIT press}
}

@article{kuppens2013relation,
  title={The relation between valence and arousal in subjective experience.},
  author={Kuppens, Peter and Tuerlinckx, Francis and Russell, James A and Barrett, Lisa Feldman},
  journal={Psychological bulletin},
  volume={139},
  number={4},
  pages={917},
  year={2013},
  publisher={American Psychological Association}
}

@article{li2018bi,
  title={A bi-hemisphere domain adversarial neural network model for EEG emotion recognition},
  author={Li, Yang and Zheng, Wenming and Zong, Yuan and Cui, Zhen and Zhang, Tong and Zhou, Xiaoyan},
  journal={IEEE Transactions on Affective Computing},
  volume={12},
  number={2},
  pages={494--504},
  year={2018},
  publisher={IEEE}
}

@article{cui2020eeg,
  title={EEG-based emotion recognition using an end-to-end regional-asymmetric convolutional neural network},
  author={Cui, Heng and Liu, Aiping and Zhang, Xu and Chen, Xiang and Wang, Kongqiao and Chen, Xun},
  journal={Knowledge-Based Systems},
  volume={205},
  pages={106243},
  year={2020},
  publisher={Elsevier}
}

@article{tao2020eeg,
  title={EEG-based emotion recognition via channel-wise attention and self attention},
  author={Tao, Wei and Li, Chang and Song, Rencheng and Cheng, Juan and Liu, Yu and Wan, Feng and Chen, Xun},
  journal={IEEE Transactions on Affective Computing},
  volume={14},
  number={1},
  pages={382--393},
  year={2020},
  publisher={IEEE}
}

@article{zhang2024mini,
  title={Mini review: Challenges in EEG emotion recognition},
  author={Zhang, Zhihui and Fort, Josep M and Gim{\'e}nez Mateu, Luis},
  journal={Frontiers in Psychology},
  year={2024}
}

@article{lima2024multimodal,
  title={Multimodal emotion classification using machine learning in immersive and non-immersive virtual reality},
  author={Lima, Rodrigo and Chirico, Alice and Varandas, Rui and Gamboa, Hugo and Gaggioli, Andrea and i Badia, Sergi Berm{\'u}dez},
  journal={Virtual Reality},
  volume={28},
  number={2},
  pages={107},
  year={2024},
  publisher={Springer}
}

@article{GramfortEtAl2013a,
  title = {{{MEG}} and {{EEG}} Data Analysis with {{MNE}}-{{Python}}},
  author = {Gramfort, Alexandre and Luessi, Martin and Larson, Eric and Engemann, Denis A. and Strohmeier, Daniel and Brodbeck, Christian and Goj, Roman and Jas, Mainak and Brooks, Teon and Parkkonen, Lauri and H{\"a}m{\"a}l{\"a}inen, Matti S.},
  year = {2013},
  volume = {7},
  pages = {1--13},
  doi = {10.3389/fnins.2013.00267},
  journal = {Frontiers in Neuroscience},
  number = {267}
}

@INPROCEEDINGS{6695876,
  author={Duan, Ruo-Nan and Zhu, Jia-Yi and Lu, Bao-Liang},
  booktitle={2013 6th International IEEE/EMBS Conference on Neural Engineering (NER)}, 
  title={Differential entropy feature for EEG-based emotion classification}, 
  year={2013},
  volume={},
  number={},
  pages={81-84},
  doi={10.1109/NER.2013.6695876}}

@article{COAN20047,
title = {Frontal EEG asymmetry as a moderator and mediator of emotion},
journal = {Biological Psychology},
volume = {67},
number = {1},
pages = {7-50},
year = {2004},
note = {Frontal EEG Asymmetry, Emotion, and Psychopathology},
issn = {0301-0511},
doi = {https://doi.org/10.1016/j.biopsycho.2004.03.002},
url = {https://www.sciencedirect.com/science/article/pii/S0301051104000316},
author = {James A Coan and John J.B Allen}
}

@article{cen2025convolution,
  title={A convolution and attention-based domain adaptation network for emotion recognition using electroencephalography},
  author={Cen, Haoming and Zhao, Mingqi and Cui, Kunbo and Tian, Fuze and Zhao, Qinglin and Hu, Bin},
  journal={Biomedical Signal Processing and Control},
  volume={100},
  pages={106957},
  year={2025},
  publisher={Elsevier}
}

@article{ghous2025attention,
  title={Attention-Driven Emotion Recognition in EEG: A Transformer-Based Approach with Cross-Dataset Fine-Tuning},
  author={Ghous, Ghulam and Najam, Shaheryar and Alshehri, Mohammed and Alshahrani, Abdulmonem and AI Qahtani, Yahya and Jalal, Ahmad and Liu, Hui},
  journal={IEEE Access},
  year={2025},
  publisher={IEEE}
}

@article{roshanaei2025eeg,
  title={EEG-based functional and effective connectivity patterns during emotional episodes using graph theoretical analysis},
  author={Roshanaei, Majid and Norouzi, Hamzeh and Onton, Julie and Makeig, Scott and Mohammadi, Alireza},
  journal={Scientific Reports},
  volume={15},
  number={1},
  pages={2174},
  year={2025},
  publisher={Nature Publishing Group UK London}
}

@article{shu2020wearable,
  title={Wearable emotion recognition using heart rate data from a smart bracelet},
  author={Shu, Lin and Yu, Yang and Chen, Wenzhuo and Hua, Haoqi and Li, Qin and Jin, Jianxiu and Xu, Xiangmin},
  journal={Sensors},
  volume={20},
  number={3},
  pages={718},
  year={2020},
  publisher={MDPI}
}

@inproceedings{khan2016recognizing,
  title={Recognizing emotion from blood volume pulse and galvanic skin response sensor using machine learning algorithms},
  author={Khan, Ali Mehmood and Lawo, Michael},
  booktitle={XIV Mediterranean Conference on Medical and Biological Engineering and Computing 2016: MEDICON 2016, March 31st-April 2nd 2016, Paphos, Cyprus},
  pages={1297--1303},
  year={2016},
  organization={Springer}
}

@article{xu2025emotion,
  title={Emotion recognition based on time-scale heterogeneity and hierarchical spatial coupling analysis of multimodal physiological signals},
  author={Xu, Zhangyong and Chen, Ning and Li, Guangqiang and Li, Jing and Zhu, Hongqing and Zhu, Zhiying},
  journal={Expert Systems with Applications},
  pages={128035},
  year={2025},
  publisher={Elsevier}
}

@inproceedings{jia2021hetemotionnet,
  title={HetEmotionNet: two-stream heterogeneous graph recurrent neural network for multi-modal emotion recognition},
  author={Jia, Ziyu and Lin, Youfang and Wang, Jing and Feng, Zhiyang and Xie, Xiangheng and Chen, Caijie},
  booktitle={Proceedings of the 29th ACM international conference on multimedia},
  pages={1047--1056},
  year={2021}
}

@article{wu2025phy,
  title={Phy-FusionNet: A Memory-Augmented Transformer for Multimodal Emotion Recognition With Periodicity and Contextual Attention},
  author={Wu, Tianyi and Purwanto, Erick and Huang, Yongrun and Yang, Su},
  journal={IEEE Transactions on Affective Computing},
  year={2025},
  publisher={IEEE}
}

@inproceedings{jiang2023multimodal,
  title={Multimodal adaptive emotion transformer with flexible modality inputs on a novel dataset with continuous labels},
  author={Jiang, Wei-Bang and Liu, Xuan-Hao and Zheng, Wei-Long and Lu, Bao-Liang},
  booktitle={proceedings of the 31st ACM international conference on multimedia},
  pages={5975--5984},
  year={2023}
}

@article{woo2025deep,
  title={Deep multimodal emotion recognition using modality-aware attention and proxy-based multimodal loss},
  author={Woo, Sungpil and Zubair, Muhammad and Lim, Sunhwan and Kim, Daeyoung},
  journal={Internet of Things},
  volume={31},
  pages={101562},
  year={2025},
  publisher={Elsevier}
}

@article{kim2025mifu,
  title={Mifu-ER: Modality Quality Index-based Incremental Fusion for Emotion Recognition},
  author={Kim, Sun-Hee},
  journal={IEEE Access},
  year={2025},
  publisher={IEEE}
}

@article{koelstra2011deap,
  title={Deap: A database for emotion analysis; using physiological signals},
  author={Koelstra, Sander and Muhl, Christian and Soleymani, Mohammad and Lee, Jong-Seok and Yazdani, Ashkan and Ebrahimi, Touradj and Pun, Thierry and Nijholt, Anton and Patras, Ioannis},
  journal={IEEE transactions on affective computing},
  volume={3},
  number={1},
  pages={18--31},
  year={2011},
  publisher={IEEE}
}

@article{katsigiannis2017dreamer,
  title={DREAMER: A database for emotion recognition through EEG and ECG signals from wireless low-cost off-the-shelf devices},
  author={Katsigiannis, Stamos and Ramzan, Naeem},
  journal={IEEE journal of biomedical and health informatics},
  volume={22},
  number={1},
  pages={98--107},
  year={2017},
  publisher={IEEE}
}

@article{miranda2018amigos,
  title={Amigos: A dataset for affect, personality and mood research on individuals and groups},
  author={Miranda-Correa, Juan Abdon and Abadi, Mojtaba Khomami and Sebe, Nicu and Patras, Ioannis},
  journal={IEEE transactions on affective computing},
  volume={12},
  number={2},
  pages={479--493},
  year={2018},
  publisher={IEEE}
}

@article{mauss2009measures,
  title={Measures of emotion: A review},
  author={Mauss, Iris B and Robinson, Michael D},
  journal={Cognition and emotion},
  volume={23},
  number={2},
  pages={209--237},
  year={2009},
  publisher={Taylor \& Francis}
}

@article{wang2018arousal,
  title={Arousal effects on pupil size, heart rate, and skin conductance in an emotional face task},
  author={Wang, Chin-An and Baird, Talia and Huang, Jeff and Coutinho, Jonathan D and Brien, Donald C and Munoz, Douglas P},
  journal={Frontiers in neurology},
  volume={9},
  pages={1029},
  year={2018},
  publisher={Frontiers Media SA}
}

@article{alarcao2017emotions,
  title={Emotions recognition using EEG signals: A survey},
  author={Alarcao, Soraia M and Fonseca, Manuel J},
  journal={IEEE transactions on affective computing},
  volume={10},
  number={3},
  pages={374--393},
  year={2017},
  publisher={IEEE}
}

@article{barrett2017theory,
  title={The theory of constructed emotion: an active inference account of interoception and categorization},
  author={Barrett, Lisa Feldman},
  journal={Social cognitive and affective neuroscience},
  volume={12},
  number={1},
  pages={1--23},
  year={2017},
  publisher={Oxford University Press}
}

@article{chicco2023matthews,
  title={The Matthews correlation coefficient (MCC) should replace the ROC AUC as the standard metric for assessing binary classification},
  author={Chicco, Davide and Jurman, Giuseppe},
  journal={BioData Mining},
  volume={16},
  number={1},
  pages={4},
  year={2023},
  publisher={Springer}
}

@article{carpenter2000bootstrap,
  title={Bootstrap confidence intervals: when, which, what? A practical guide for medical statisticians},
  author={Carpenter, James and Bithell, John},
  journal={Statistics in medicine},
  volume={19},
  number={9},
  pages={1141--1164},
  year={2000},
  publisher={Wiley Online Library}
}

@article{schechtman2010negative,
  title={Negative valence widens generalization of learning},
  author={Schechtman, Eitan and Laufer, Offir and Paz, Rony},
  journal={Journal of Neuroscience},
  volume={30},
  number={31},
  pages={10460--10464},
  year={2010},
  publisher={Society for Neuroscience}
}

@article{zong2025accuracy,
  title={Accuracy, Memory Efficiency and Generalization: A Comparative Study on Liquid Neural Networks and Recurrent Neural Networks},
  author={Zong, Shilong and Bierly, Alex and Boker, Almuatazbellah and Eldardiry, Hoda},
  journal={arXiv preprint arXiv:2510.07578},
  year={2025}
}

@inproceedings{hasani2021liquid,
  title={Liquid time-constant networks},
  author={Hasani, Ramin and Lechner, Mathias and Amini, Alexander and Rus, Daniela and Grosu, Radu},
  booktitle={Proceedings of the AAAI Conference on Artificial Intelligence},
  volume={35},
  pages={7657--7666},
  year={2021}
}

@inproceedings{glorot2010understanding,
  title={Understanding the difficulty of training deep feedforward neural networks},
  author={Glorot, Xavier and Bengio, Yoshua},
  booktitle={Proceedings of the thirteenth international conference on artificial intelligence and statistics},
  pages={249--256},
  year={2010},
  organization={JMLR Workshop and Conference Proceedings}
}

@article{srivastava2014dropout,
  title={Dropout: a simple way to prevent neural networks from overfitting},
  author={Srivastava, Nitish and Hinton, Geoffrey and Krizhevsky, Alex and Sutskever, Ilya and Salakhutdinov, Ruslan},
  journal={The journal of machine learning research},
  volume={15},
  number={1},
  pages={1929--1958},
  year={2014},
  publisher={JMLR. org}
}

@article{zheng2015investigating,
  title={Investigating Critical Frequency Bands and Channels for {EEG}-based Emotion Recognition with Deep Neural Networks},
  author={Zheng, Wei-Long and Lu, Bao-Liang},
  journal={IEEE Transactions on Autonomous Mental Development},
  doi={10.1109/TAMD.2015.2431497},
  year={2015},
  volume={7},
  number={3},
  pages={162-175},
  publisher={IEEE}
}

@article{van2008visualizing,
  title={Visualizing data using t-SNE.},
  author={Van der Maaten, Laurens and Hinton, Geoffrey},
  journal={Journal of machine learning research},
  volume={9},
  number={11},
  year={2008}
}

@article{calinski1974dendrite,
  title={A dendrite method for cluster analysis},
  author={Cali{\'n}ski, Tadeusz and Harabasz, Jerzy},
  journal={Communications in Statistics-theory and Methods},
  volume={3},
  number={1},
  pages={1--27},
  year={1974},
  publisher={Taylor \& Francis}
}

@ARTICLE{4766909,
  author={Davies, David L. and Bouldin, Donald W.},
  journal={IEEE Transactions on Pattern Analysis and Machine Intelligence}, 
  title={A Cluster Separation Measure}, 
  year={1979},
  volume={PAMI-1},
  number={2},
  pages={224-227}
}

@article{mahalanobis2018generalized,
  title={On the generalized distance in statistics},
  author={Mahalanobis, Prasanta Chandra},
  journal={Sankhy{\=a}: The Indian Journal of Statistics, Series A (2008-)},
  volume={80},
  pages={S1--S7},
  year={2018},
  publisher={JSTOR}
}

@article{loshchilov2017decoupled,
  title={Decoupled weight decay regularization},
  author={Loshchilov, Ilya and Hutter, Frank},
  journal={arXiv preprint arXiv:1711.05101},
  year={2017}
}

@misc{loshchilov2017sgdrstochasticgradientdescent,
      title={SGDR: Stochastic Gradient Descent with Warm Restarts}, 
      author={Ilya Loshchilov and Frank Hutter},
      year={2017},
      eprint={1608.03983},
      archivePrefix={arXiv},
      primaryClass={cs.LG},
      url={https://arxiv.org/abs/1608.03983}, 
}

@article{OLOFSSON2008247,
title = {Affective picture processing: An integrative review of ERP findings},
journal = {Biological Psychology},
volume = {77},
number = {3},
pages = {247-265},
year = {2008},
issn = {0301-0511},
doi = {https://doi.org/10.1016/j.biopsycho.2007.11.006},
url = {https://www.sciencedirect.com/science/article/pii/S0301051107001913},
author = {Jonas K. Olofsson and Steven Nordin and Henrique Sequeira and John Polich}
}

@article{Hajcak12022010,
author = {Greg Hajcak and Annmarie MacNamara and Doreen M. Olvet},
title = {Event-Related Potentials, Emotion, and Emotion Regulation: An Integrative Review},
journal = {Developmental Neuropsychology},
volume = {35},
number = {2},
pages = {129--155},
year = {2010},
publisher = {Routledge},
doi = {10.1080/87565640903526504},
note ={PMID: 20390599},
URL = {https://doi.org/10.1080/87565640903526504},
eprint = {https://doi.org/10.1080/87565640903526504}
}

@article{10.1093/biomet/52.3-4.591,
    author = {SHAPIRO, S. S. and WILK, M. B.},
    title = {An analysis of variance test for normality (complete samples)†},
    journal = {Biometrika},
    volume = {52},
    number = {3-4},
    pages = {591-611},
    year = {1965},
    month = {12},
    issn = {0006-3444},
    doi = {10.1093/biomet/52.3-4.591},
    url = {https://doi.org/10.1093/biomet/52.3-4.591},
    eprint = {https://academic.oup.com/biomet/article-pdf/52/3-4/591/962907/52-3-4-591.pdf},
}

@article{8ddb7f85-9a8c-3829-b04e-0476a67eb0fd,
 ISSN = {00359254, 14679876},
 URL = {http://www.jstor.org/stable/2346830},
 author = {J. A. Hartigan and M. A. Wong},
 journal = {Journal of the Royal Statistical Society. Series C (Applied Statistics)},
 number = {1},
 pages = {100--108},
 publisher = {[Royal Statistical Society, Oxford University Press]},
 title = {Algorithm AS 136: A K-Means Clustering Algorithm},
 urldate = {2026-01-24},
 volume = {28},
 year = {1979}
}

@article{Spearman_1904, title={The proof and measurement of association between two things}, 
volume={15}, 
DOI={10.2307/1412159}, 
number={1}, 
journal={The American Journal of Psychology}, 
author={Spearman, C.}, 
year={1904}, 
month={Jan}, 
pages={72}
}

@InProceedings{pmlr-v28-pascanu13,
  title = 	 {On the difficulty of training recurrent neural networks},
  author = 	 {Pascanu, Razvan and Mikolov, Tomas and Bengio, Yoshua},
  booktitle = 	 {Proceedings of the 30th International Conference on Machine Learning},
  pages = 	 {1310--1318},
  year = 	 {2013},
  editor = 	 {Dasgupta, Sanjoy and McAllester, David},
  volume = 	 {28},
  number =       {3},
  series = 	 {Proceedings of Machine Learning Research},
  address = 	 {Atlanta, Georgia, USA},
  month = 	 {17--19 Jun},
  publisher =    {PMLR},
  url = 	 {https://proceedings.mlr.press/v28/pascanu13.html},
}

@inproceedings{guo2017calibration,
  title={On calibration of modern neural networks},
  author={Guo, Chuan and Pleiss, Geoff and Sun, Yu and Weinberger, Kilian Q},
  booktitle={International conference on machine learning},
  pages={1321--1330},
  year={2017},
  organization={PMLR}
}

@article{glenn1950verification,
  title={Verification of forecasts expressed in terms of probability},
  author={Glenn, W Brier and others},
  journal={Monthly weather review},
  volume={78},
  number={1},
  pages={1--3},
  year={1950},
  publisher={War Department, Office of the Chief Signal Officer}
}

@article{platt1999probabilistic,
  title={Probabilistic outputs for support vector machines and comparisons to regularized likelihood methods},
  author={Platt, John and others},
  journal={Advances in large margin classifiers},
  volume={10},
  number={3},
  pages={61--74},
  year={1999},
  publisher={Cambridge, MA}
}

@article{hochreiter1997long,
  title={Long short-term memory},
  author={Hochreiter, Sepp and Schmidhuber, J{\"u}rgen},
  journal={Neural computation},
  volume={9},
  number={8},
  pages={1735--1780},
  year={1997},
  publisher={MIT press}
}

@article{cho2014learning,
  title={Learning phrase representations using RNN encoder-decoder for statistical machine translation},
  author={Cho, Kyunghyun and Van Merri{\"e}nboer, Bart and Gulcehre, Caglar and Bahdanau, Dzmitry and Bougares, Fethi and Schwenk, Holger and Bengio, Yoshua},
  journal={arXiv preprint arXiv:1406.1078},
  year={2014}
}

@article{bai2018empirical,
  title={An empirical evaluation of generic convolutional and recurrent networks for sequence modeling. arXiv},
  author={Bai, Shaojie and Kolter, J Zico and Koltun, Vladlen},
  journal={arXiv preprint arXiv:1803.01271},
  volume={10},
  year={2018}
}

@article{chen2018neural,
  title={Neural ordinary differential equations},
  author={Chen, Ricky TQ and Rubanova, Yulia and Bettencourt, Jesse and Duvenaud, David K},
  journal={Advances in neural information processing systems},
  volume={31},
  year={2018}
}

@article{Khanh_Kim_Lee_Yang_Baek_2020,
title={Korean video dataset for emotion recognition in the wild}, 
volume={80}, 
DOI={10.1007/s11042-020-10106-1}, 
number={6}, 
journal={Multimedia Tools and Applications}, 
author={Khanh, Trinh Le and Kim, Soo-Hyung and Lee, Gueesang and Yang, Hyung-Jeong and Baek, Eu-Tteum}, 
year={2020}, 
month={Nov}, 
pages={9479–9492}
}

\end{document}